\documentclass[prd,aps,preprint,tightenlines,superscriptaddress]{revtex4}
\usepackage{epsfig}
\usepackage{amsmath}
\usepackage{axodraw}
\preprint{DOE/ER/40762-280}
\preprint{UM-PP\#03-046} 

\def \p {\partial}

\def \pn3 {\psi_{u\bar d g}}

\def \pu4 {\psi_{u\bar d q\bar q}}

\def \p4n {\psi_{u\bar d gg}}

\def \pp4 {\psi_{uudg}}

\begin{document}
\title{Classification and Asymptotic Scaling of \\
Hadrons' Light-Cone Wave-Function Amplitudes}
\author{Xiangdong Ji}
\email{xji@physics.umd.edu}
\affiliation{Department of Physics,
University of Maryland,
College Park, Maryland 20742, USA}
\author{Jian-Ping Ma}
\email{majp@itp.ac.cn}
\affiliation{Institute of Theoretical Physics,
Academia Sinica, Beijing, 100080, P. R. China}
\author{Feng Yuan}
\email{fyuan@physics.umd.edu}
\affiliation{Department of Physics,
University of Maryland,
College Park, Maryland 20742, USA}

\date{\today}
\vspace{0.5in}          
\begin{abstract}
We classify the hadron light-cone wave-function amplitudes
in terms of parton helicity, orbital angular momentum, and quark-flavor
and color symmetries. We show in detail how this is done for
the pion, $\rho$ meson, nucleon, and delta resonance 
up to and including three partons. For the pion and nucleon, we
also consider four-parton amplitudes. Using the scaling law
derived previously, we show how these amplitudes scale in the limit 
that all parton transverse momenta become large.
\end{abstract}

\maketitle

\section{introduction}
Although the hadron structure is believed to be described
by the fundamental theory of strong interactions---quantum 
chromodynamics (QCD), the actual solution of the problem 
is notoriously difficult to achieve. Apart from numerically 
solving QCD on a spacetime lattice, there is no other systematic 
theoretical approach that has been very successful. The closest
might be the light-front quantization approach in which the 
old-fashioned method of diagonalizing a hamiltonian is followed
\cite{Brodsky:1998de,Burkardt:1996ct}. 
The conceptual advantage here is obvious: Hadrons are 
described by light-cone wave functions which have clear physical 
meaning and are very useful phenomenologically, whereas in lattice QCD 
the natural language is classical gluon configurations, such as 
instantons and monopoles, in the Euclidean space. As to why 
light-cone quantization is superior 
compared to equal-time quantization, we just wish to point out that 
the vacuum structure in the former approach, which consists of just 
$k^+=0$ particles, can be easily separated from the part of the 
hadron structure consisting $k^+\ne 0$ particles. Moreover, in high-energy
scattering hadrons travel at the near speed of light, and the
light-cone coordinates appear naturally. 

To be sure, there are many difficulties that one must clear 
before a realistic light-cone description of hadrons becomes possible.  
One of them is that hadrons are now described by an infinite number
of light-cone Fock amplitudes, and there is no apparent reason why 
the amplitudes with 100 partons (quarks and gluons) are strongly 
suppressed relative to those with 2 or 3 partons. The answer, 
of course, depends on the choice of gauge, ultraviolet and 
infrared cut-offs, and ultimately the underlying QCD dynamics. 
One way to check is to truncate the Hilbert space first to 
including the partons to a maximum number $n$ and then to determine 
how the solution changes when the Fock components with $n+1$ number 
of partons are included. The optimistic view has been that 
since the constituent quark models work so well phenomenologically, 
there must exist a light-cone
description of hadrons in which only the Fock components 
with a few partons are necessary. In high-energy exclusive processes, 
we know for sure that only the wave-function components 
with a few partons are relevant. 

In the light-front description of a hadron, the very first step 
is to classify independent wave-function amplitudes
given a particular parton combination. To our knowledge, there has not been 
much systematic study in the literature along this direction. 
In Ref. \cite{Burkardt:2002uc}, we have proposed an approach by writing 
down the matrix elements of a class of light-cone-correlated 
quark-gluon operators, in much the same way that has been used to 
construct independent light-cone distribution amplitudes in which 
the parton transverse momenta are integrated out
\cite{Braun:1999te}. In Ref. \cite{Ji:2002xn}, we have applied 
the approach to the nucleon, finding that six 
amplitudes are needed to describe the three-quark sector of the 
nucleon wave function. However, using the approach 
to handle Fock states with more partons appears to be complicated. 

In \cite{Ji:2003fw}, we have developed a more direct method to 
write down the general structure of the light-cone
wave function for $n$ partons with orbital angular momentum 
projection $l_z$. We have also found a general power counting rule 
which determines the asymptotic behavior of the light-cone 
amplitudes when the transverse momenta of all partons become large. 
From the wave-function counting
rule, we have derived the dimensional scaling law for high-energy 
exclusive processes including parton orbital angular momentum 
\cite{Brodsky:1973kr,Matveev:1973ra}. 
In this paper, we report further progress in 
this direction. In particular, we use the method to 
classify the higher Fock components of hadrons. We will consider in 
detail how the flavor (for quarks) and color degrees of freedom 
of the partons are systematically coupled. We will write down the 
amplitudes for the pion and proton up to four partons. 
We also work out the leading light-cone wave functions for 
the $\Delta$ and $\rho$ meson, leaving more complicated cases for
the readers. 

Based on our work here, one can go on to parameterize the light-cone 
wave function amplitudes and fit to the experimental data. Although the
amplitudes thus determined are phenomenological, they can be made to
obey the asymptotic behavior at large transverse momenta \cite{Ji:2003fw}.
Therefore, our result can provide guidelines for phenomenological studies 
for exclusive processes \cite{Chung:1991st,Schlumpf:1993vq,Bolz:1996sw,
Diehl:1998kh}. 
By committing ourselves to the light-cone amplitudes, 
we are also committing to the light-cone gauges $A^+=0$
\cite{Kogut:1970xa,Lepage:1980fj}. 
One subtlety about the light-cone gauge is that it requires 
additional gauge fixing \cite{Mueller:1986wy,Slavnov:1987yh,Antonov:1988yg}. 
Depending on whether the additional gauge condition is time-reversal
invariant or not, the wave function amplitudes are real or fully complex. 
In the latter case, the final state interaction effects
might be included in the amplitudes 
\cite{Kovchegov:1997pc,Ji:2002aa,Belitsky:2002sm,Boer:2003cm}.
A related issue is that the light-cone amplitudes have light-cone
singularities at small $x$ which require regularization. 

Our plan of the presentation is as follows. We start in Sec. II by 
describing a general strategy to classify the independent
wave-function amplitudes for a hadron state with a specific
parton content.
In Sec. III, we apply this method to write down the 
amplitudes of $\pi^+$
for up to four-parton Fock components. 
We extend these discussions to $\rho$ mesons in Sec. IV, where
the amplitudes up to three-parton Fock components
will be given.
In Sec. V, the proton wave-function amplitudes 
for three-quark and three-quark plus one-gluon Fock components
will be presented. The leading results for the delta resonance 
will be given in Sec. VI. The final section contain a brief summary
and outlook. 

\section{general Strategy and Symmetry Constraints}

In this section, we discuss our general strategy in classifying
and enumerating the hadron light-cone amplitudes. The goal is to
find a simple and general way to write done all possible 
light-cone amplitudes of a hadron once a parton content is specified. 
In Sec. II.A, we explain our notation and convections. In 
Sec. II.B, we consider the helicity and angular momentum structure of a general
Fock component. In Sec. II.C we make general comments about
flavor and color structure. In Sec. II.D and E, we consider 
the parity and time reverse constraints on
the light-cone wave function amplitudes.

\subsection{Notation}

We work in the framework of light-cone (or light-front) quantization
\cite{Lepage:1980fj,Brodsky:1998de}.
The light-cone time $x^+$ and coordinate $x^-$ are defined as $x^\pm
=1/\sqrt{2}(x^0\pm x^3)$. Likewise we define Dirac matrices
$\gamma^\pm = 1/\sqrt{2}(\gamma^0\pm\gamma^3)$. The 
projection operators for Dirac fields are defined 
as $P_\pm = (1/2)\gamma^\mp\gamma^\pm$. Any Dirac field $\psi$ 
can be decomposed into $\psi=\psi_++\psi_-$ with
$\psi_\pm = P_\pm \psi$. $\psi_+$ is a dynamical degrees of freedom 
and has the canonical expansion, 
\begin{eqnarray}
  \psi_+(\xi^+=0,\xi^-,\xi_\perp)
  &=& \int \frac{d^2k_\perp}{ (2\pi)^3}
    \frac{dk^+}{ 2k^+}
   \sum_\lambda
  \left[b_\lambda(k) u(k\lambda) e^{-i(k^+\xi^--\vec{k}_\perp\cdot\vec{\xi}_\perp)} \right. \nonumber \\
 && \left. + d_\lambda^\dagger(k) v(k\lambda)e^{i(k^+\xi^--\vec{k}_\perp\cdot\vec{\xi}_\perp)}
 \right]
  \ , 
\end{eqnarray}
where $b^\dagger(b)$ and $d^\dagger(d)$ are quark and antiquark creation (annihilation)
operators, respectively. We adopt covariant normalization for the particle
states and the creation and annihilation operators, i.e.,
\begin{equation}
\big\{b_\lambda(k),b_{\lambda'}^\dagger(k')\big\}=
\big\{d_\lambda(k),d_{\lambda'}^\dagger(k')\big\}=
(2\pi)^3\delta_{\lambda\lambda'}2k^+\delta(k^+-k^{'+})
\delta^{(2)}(\vec{k}_\perp-\vec{k}_\perp^\prime)\ ,
\end{equation}
where $\lambda$ is the light-cone
helicity of the quarks which can take $+1/2$ or $-1/2$. We ignore 
the masses of the light up and down quarks. Later, to simply the 
notations, we simply use $u^\dagger$ and $\overline{u}^\dagger$ to represent 
creation operators for up and anti-up quarks, respectively, and so on.  

Likewise, for the gluon fields in the light-cone gauge $A^+=0$, 
$A_\perp$ is dynamical and has the expansion,
\begin{eqnarray}
    A_\perp(\xi^+=0,\xi^-,\xi_\perp)
  &=&  \int \frac{d^2k_\perp}{(2\pi)^3}
    \frac{dk^+}{ 2k^+}
   \sum_\lambda
  \left[ a_\lambda(k) \epsilon(k\lambda) e^{-i(k^+\xi^--\vec{k}_\perp\cdot\vec{\xi}_\perp)}
   \right.
  \nonumber \\
  && \left. + a_\lambda^\dagger(k) \epsilon^*(k\lambda)e^{i(k^+\xi^--\vec{k}_\perp
   \cdot \vec{\xi}_\perp)} \right] \ .
\end{eqnarray}
Implicitly, the gauge fields $A^\mu$ is a traceless $3\times 3$ 
matrix with $A^\mu=\sum_aA^{a\mu}T^a$, where $T^a$ are the
$SU(3)$ Gell-Mann matrices satisfying $[T^a,T^b]=if^{abc}T^c$
and $\{T^a,T^b\}=\frac{1}{3}\delta_{ab}+d_{abc}T^c$.
Again, we have the following covariant normalization for the 
creation and annihilation operators for gluon,
\begin{equation}
\left[a_\lambda(k),a_{\lambda'}^\dagger(k')\right]=(2\pi)^3\delta_{\lambda\lambda'}2k^+\delta(k^+-k^{'+})
\delta^{(2)}(\vec{k}_\perp-\vec{k}_\perp^\prime)\ ,
\end{equation}
Later, we simply use $g^\dagger$ to
represent a gluon creation operator. 
$\psi_-$ and $A^-$ are dependent variables, which can be expressed in 
terms of $\psi_+$ and $A_\perp$ using equations of motion \cite{Kogut:1970xa}.

\subsection{Angular Momentum Structure}

For a given parton content, i.e., a specification of quarks, antiquarks
and gluons, the light-cone amplitudes of a hadron with helicity $\Lambda$ 
can be classified in terms of the total parton light-cone helicity $\lambda$. 
The angular momentum conservation then demands that the partons have 
angular momentum projection $l_z=\Lambda-\lambda$. Let us find the 
angular momentum structure of the amplitudes satisfying these conditions
~\cite{Ji:2003fw}. 

Suppose a Fock component 
has $n$ partons with creation operators $a^\dagger_1$, 
..., $a^\dagger_{n}$, where the partons can either be gluons or 
quarks and the subscripts label the partons' quantum numbers such as
spin, flavor, color, momentum, etc. Assume all color, 
flavor (for quarks) indices have been coupled properly using 
Clebsch-Gordon coefficients (see next subsection). 
The longitudinal momentum fractions of the partons
are $x_i$ $(i=1,2,...,n)$, satisfying $\sum_{i=1}^n x_i = 1$, 
and the transverse momenta $\vec{k}_{1\perp},...,\vec{k}_{n\perp}$, satisfying
$\sum_{i}^n\vec{k}_{i\perp}=0$. We will eliminate $\vec{k}_{n\perp}$ in favor 
of the first $n-1$ transverse momenta. Assume the orbital angular 
momentum projections of the partons are $l_{z1},...,l_{z(n-1)}$, 
respectively, and let $l_z = \sum_{i=1}^{n-1} l_{zi}$, then
\begin{equation}
          l_z + \lambda = \Lambda \ , 
\end{equation}
where $\lambda=\sum_{i=1}^n\lambda_i$ is the total parton helicity.
Without loss of generality, we assume $l_z\ge 0$; even then, $l_{zi}$
can have both signs. Thus, a general term in the hadron wave function 
amplitude has the structure
\begin{eqnarray}
     \int \prod_{i=1}^n d[i]
        (k_{1\perp}^\pm)^{|l_{z1}|} (k_{2\perp}^\pm)^{|l_{z2}|} 
            ... (k_{(n-1)\perp}^\pm)^{|l_{z(n-1)}|}
	\psi_n(x_i,k_{i\perp},\lambda_i,l_{z i})~
    a_1^\dagger a_2^\dagger ... a_n^\dagger  |0\rangle \ , 
\end{eqnarray}
where $k_{i\perp}^\pm = k_{i}^x \pm k_{i}^y$ and the $+(-)$ sign applies
when $l_{zi}$ is positive (negative), and 
$$d[i] = dx_id^2k_{i\perp}/(\sqrt{2x_i}(2\pi)^3)$$ 
with the overall constraint on $x_i$ and $k_{i\perp}$ implicit.

The above form can be further simplified as follows. Assume $l_{zi}$ is positive 
and $l_{zj}$ negative, and $l_{zi}>|l_{zj}|$, we have
\begin{eqnarray}
   (k_{i\perp}^+)^{l_{zi}}(k_{j\perp}^-)^{-l_{zj}}
	&=&(k_{i\perp}^+)^{l_{zi}+l_{zj}}(k_{i\perp}^+k_{j\perp}^-)^{-l_{zj}} \nonumber \\
   &=& (k_i^+)^{l_{zi}+l_{zj}}(\vec{k}_{i\perp}\cdot \vec{k}_{j\perp}-i\epsilon^{\alpha\beta}
          k_{i\alpha}k_{j\beta})^{-l_{zj}} \nonumber \\
   &=& (k_{i\perp}^+)^{l_{zi}+l_{zj}}(\phi_0 + \phi_1 
               i\epsilon^{\alpha\beta}
          k_{i\alpha}k_{j\beta}) \ ,\nonumber 
\end{eqnarray}
where $\alpha,\beta=1,2$, $\phi_{0,1}$ are polynomials in $\vec{k}^2_{i\perp}$, 
$\vec{k}^2_{j\perp}$, and $\vec{k}_{i\perp}\cdot \vec{k}_{j\perp}$. On the last line of 
the above equation we have used the identity 
$ \epsilon^{\alpha\beta} \epsilon^{\gamma\delta}
 = \delta^{\alpha\gamma} \delta^{\beta\delta} - \delta^{\alpha\delta} 
       \delta^{\beta\gamma}$. 
If $l_{zi}+l_{zj}\ne 0$, one can use 
$i\epsilon^{\alpha\beta} k_{1\alpha} k_{2\beta} k_{1\perp}^+
         = \vec{k}_{1\perp}\cdot \vec{k}_{2\perp} k_{1\perp}^+ 
        - \vec{k}_{1\perp}\cdot \vec{k}_{1\perp} k_{2\perp}^+$
to further reduce the second term in the bracket.
Following the above procedure, we can eliminate all negative 
$l_{zj}$, a general $l_z>0$ component in the wave function reads
\begin{eqnarray}
   && \int \prod_{i=1}^n d[i]~~
        (k_{1\perp}^+)^{l_{z1}} (k_{2\perp}^+)^{l_{z2}} 
            ... (k_{(n-1)\perp}^+)^{l_{z(n-1)}}~ \nonumber \\ 
 &&  ~~~~ \times \left(\psi_{n}(x_i,k_i,\lambda_i,l_{zi})
          + \sum_{i<j=1|_{ l_{zi}=l_{zj}=0}}^{n-1} i\epsilon^{\alpha\beta} 
              k_{i\alpha}k_{j\beta}\psi_{n(ij)}(x_i,k_{i\perp},\lambda_i,l_{zi})\right)~
    a_1^\dagger a_2^\dagger ... a_n^\dagger  |0\rangle 
\label{gs}
\end{eqnarray}
where $\sum_i l_{zi}=l_z$ and $l_{zi}\ge 0$, and the sums over $i$ and 
$j$ are restricted to the $l_{zi}=0$ partons. The above equation is
our starting point to write down independent light-cone
amplitudes. 

\subsection{Flavor and Color Structure}

For a given quark content, we classify the amplitudes 
in terms of the flavor symmetry. For instance, for the
pion state, we need to project out the Fock component 
with the total isospin 1. The problem becomes more involved 
if a Fock state contains many quark-antiquark pairs because
there are more than one way to construct the states with the
definite isospin. 

Our general strategy is as follows: we first consider 
all possible ways to construct the same isospin. We then use 
the freedom that the labels of the quark partons are
arbitrary to shuffle the particles around. If after the shuffling,
the flavor content of a combination is identical to the one considered before,
the combination is ignored. For example, consider $u\bar d q\bar q$
component of a $\pi^+$ particle. The $q\bar q$ 
can either couple to $I=1$ or $I=0$. It turns out that the
combination coupled to $I=1$ is not independent after reshuffling
of the particle label.

All the hadrons are color neutral. Therefore, we couple all partons
to color singlet. All possible ways of making the coupling must
be considered.

\subsection{Parity}

Consider a hadron moving in the $z$-direction with helicity $\Lambda$, $|P\Lambda\rangle$.
Under parity transformation, the momentum changes the direction, and the helicity changes
sign. However, if we make an additional $180^\circ$ rotation around the $y$ axis, the original 
momentum is restored, and we have a state $|P-\Lambda\rangle$. According to Jacob
and Wick \cite{Jacob:1959at}, we have, 
\begin{equation}
       (-1)^{s-\Lambda}\eta|P-\Lambda\rangle = \hat Y|P\Lambda\rangle
\label{jw}
\end{equation}
where $\hat Y$ is a parity operation followed by a $180^\circ$ rotation around 
the $y$ axis, and $\eta$ is the intrinsic parity of the hadron. 

For a particle state with non-zero helicity, the above equation allows one to obtain
the wave function of the state with helicity $(-\Lambda)$ from that
with helicity $\Lambda$. On the other hand, for a particle of zero helicity, the
above equation can be considered as a constraint on the wave function. 
 
When $\hat Y$ acts on the individual partons, the transformation is
\begin{equation}
  (-1)^{s-\lambda}\eta|k_x,-k_y,k_z,-\lambda\rangle 
           = \hat Y|k_x,k_y,k_z,\lambda\rangle
\end{equation}
where the intrinsic parity for a quark is $+1$, an antiquark $-1$, and a gluon $-1$. 
For instance, for a $u$ quark state, 
\begin{equation}
\hat Y \vert u_\uparrow \rangle = \vert u_\downarrow \rangle, \ \ \ \
\hat Y \vert u_\downarrow \rangle = -\vert u_\uparrow \rangle,
\end{equation}
where we have omitted the momentum label. For a $\bar d$ quark state,
\begin{equation}
\hat Y \vert \bar d_\uparrow \rangle =- \vert \bar d_\downarrow \rangle, \ \ \ \
\hat Y \vert \bar d_\downarrow \rangle = \vert \bar d_\uparrow \rangle,
\end{equation}
because of the opposite intrinsic parity.

\subsection{Time Reversal}

Time reversal usually provides constraint on reality 
of the wave function amplitudes. Under the transformation, however,
the light-cone time and coordinate interchange. To preserve the original
light-cone coordinates, we consider the combined time-reversal and parity. 

The light-cone gauge condition is invariant under the combined transformation. 
However, $A^+=0$ does not fix the gauge freedom completely, additional 
gauge fixing must be specified. Physically the additional gauge fixing 
corresponds to a choice of boundary conditions for gauge fields 
at $\xi^-=\pm \infty$. If one chooses the antisymmetric boundary condition,
$A_\perp(\xi^-=-\infty) = -A_\perp(\xi^-=\infty)$, which is invariant under
the combined transformation, then one can show that all light-cone
wave function amplitudes are real (principal-value prescription). On the other hand, 
if one chooses either advanced or retarded boundary conditions $A_\perp (\xi^-=\pm \infty)=0$, 
the combined transformation is broken, and the wave function amplitudes 
are complex. 

\newpage

\section{Wave-function Amplitudes for the pion}

In this section, we classify the light-cone wave-function amplitudes
for the $\pi^+$ meson up to and including four partons. 
The amplitudes for other isotriplet members can be obtained 
by using the isospin lowering operator. The pion 
is a pseudoscalar meson with spin $J=0$ and parity $P=-1$.
These quantum numbers are necessary constraints when 
the light-cone wave function amplitudes are constructed.

A pion moving in the $z$ direction has the following 
transformation under $\hat Y$,
\begin{equation}
\hat Y \vert \pi^+ \rangle =-\vert \pi^+ \rangle \ .
\label{piy}
\end{equation}
Every Fock component we write down must have this symmetry. 

In the following subsections, we present the 
wave-function amplitudes of $\pi^+$ up to four-particle
component Fock states, i.e., $u\overline d$, $u\overline d g$, 
$u\overline d gg$, and $u\overline d q\overline q$.
Related studies on the light-cone distribution amplitudes for
$\pi$ mesons can be found in 
\cite{Lepage:1979zb,Lepage:1980fj,Chernyak:1984ej,Braun:1990iv}.

\subsection{The $u\bar d$ Component}

For this component, $n=2$, and the total quark helicity
$\lambda$ can be either $0$ or $1$. The isospin does not provide any
additional constraint. From Eq. (\ref{gs}) we can have two 
wave-function amplitudes, corresponding to $l_z=0$ and $|l_z|=1$.
They have been discussed in \cite{Burkardt:2002uc} and many other
references before. For completeness, we present 
the results here,
\begin{eqnarray}
  |\pi^+\rangle^{l_z=0}_{u\overline d} &=& \int d[1]d[2] \psi_{u\overline d}^{(1)}(1,2)  
\frac{\delta_{ij}}{\sqrt{3}}\left[u^\dagger_{\uparrow i}(1)
  \overline d^\dagger_{\downarrow j}(2) - 
u^\dagger_{\downarrow i}(1)
\overline d^\dagger_{\uparrow j}(2)\right]
 |0\rangle \, \\ 
  |\pi^+\rangle _{u\overline d}^{|l_z|=1}&=& \int d[1]d[2] \psi_{u\overline d}^{(2)}(1,2)  
\frac{\delta_{ij}}{\sqrt{3}}\left[ k_{1\perp}^-u^\dagger_{\uparrow i}(1)
  \overline d^\dagger_{\uparrow j}(2)
+ k_{1\perp}^+u^\dagger_{\downarrow i}(1)
  \overline d^\dagger_{\downarrow j}(2)\right] |0\rangle \ , 
\end{eqnarray}
where $i$ and $j=1,2,3$ are the color indices, and $\uparrow$ and $\downarrow$ label
quark light-cone helicity $+1/2$ and $-1/2$, respectively.
The color factor $\delta_{ij}/\sqrt{3}$ is normalized to 1. 
The amplitudes $\psi_{u\bar d}^{(1,2)}(1,2)$ are functions of 
quark momenta with argument 1
representing $x_1$ and $k_{1\perp}$ and so on. 
The dependence on the transverse momenta is of form 
$\vec{k}_{i\perp}\cdot\vec{k}_{j\perp}$ only. 
Since the momentum conservation implies $\vec{k}_{1\perp}+\vec{k}_{2\perp}=0$
and $x_1+x_2=1$,
$\psi_{u\bar d}^{(1,2)}(1,2)$ depend on
variables $x_1$ and $k_{1\perp}^2$ only. The 
integration in the above equation become,
	$$\int d[1]d[2]=\int \frac{d^2k_{1\perp}}{(2\pi)^3}\frac{dx_1}{2\sqrt{x_1(1-x_1)}} \ . $$
It is easy to check that the amplitudes $\psi_{u\bar d}^{(1,2)}(1,2)$
have the correct transformation behavior under $\hat Y$.

\subsection{The $u\bar d g$ Component}

For this component, $n=3$, and total parton helicity $\lambda$ 
can be 0, 1, or 2. Therefore, the light-cone wave function
amplitudes must have $|l_z|=$0, 1, or 2. Again isospin
symmetry does not provide any constraint.

To satisfy the
constraint from parity, we consider the $u\bar d$ pair with 
definite properties under the $\hat Y$ transformation
\begin{eqnarray}
(u\overline d)^\dagger_{S,0}&=& u^\dagger_{\uparrow i}(1)\overline d^\dagger_{\downarrow j}(2) 
+ u^\dagger_{\downarrow i}(1) \overline d^\dagger_{\uparrow j}(2) \ ,
\nonumber\\
(u\overline d)^\dagger_{A,0}&=& u^\dagger_{\uparrow i}(1)\overline d^\dagger_{\downarrow j}(2) 
- u^\dagger_{\downarrow i}(1) \overline d^\dagger_{\uparrow j}(2) \ ,
\nonumber\\
(u\overline d)^\dagger_{A,1}&=& u^\dagger_{\uparrow i}(1)\overline d^\dagger_{\uparrow j}(2) 
\ ,\nonumber\\
(u\overline d)^\dagger_{A,-1}&=& u^\dagger_{\downarrow i}(1)\overline d^\dagger_{\downarrow j}(2) 
\ .
\end{eqnarray}
It is clear that
\begin{equation}
\hat Y (u\overline d)^\dagger_{S,\lambda_z}|0\rangle=(u\overline d)^\dagger_{S,-\lambda_z}|0\rangle, \ \ \ \ 
\hat Y (u\overline d)^\dagger_{A,\lambda_z}|0\rangle=-(u\overline d)^\dagger_{A,-\lambda_z}|0\rangle \ .
\end{equation}
where we have neglected the transformation of the momentum labels. On the other hand,
the one-gluon state transforms under $\hat Y$
\begin{equation}
\hat Y \vert g_{\lambda} \rangle = -\vert g_{-\lambda}\rangle \ .
\end{equation}
because the gluon is a vector particle. 

There is only one way to couple the color indices. The quark and anti-quark
(with color indices $i$ and $j$)
couple to an octet which in turn couples to the octet gluon 
(with color index $a$) to yield
a singlet. The coupling can be achieved with an SU(3) matrices $T^a_{ij}$. 

When $l_z=0$, the helicity of the quarks must be $\lambda_{u\bar d}=\pm 1$ because
$\lambda_g=\mp 1$. From Eq. (\ref{gs}) we have two independent amplitudes,
\begin{eqnarray}
  |\pi^+\rangle^{l_z=0}_{u\overline d g} &=& \int d[1]d[2]d[3] 
\frac{T^a_{ij}}{2}
\left\{\psi_{u\overline d g}^{(1)}(1,2,3)
\left[(u\overline d)^\dagger_{A,1}g^{a\dagger}_\downarrow (3)-(u\overline d)^\dagger_{A,-1}g^{a\dagger}_\uparrow (3)\right]
\right.\nonumber\\
&&\left.+i\epsilon^{\alpha\beta}k_{1\alpha}k_{2\beta}\psi_{u\overline d g}^{(2)}(1,2,3)
\left[(u\overline d)^\dagger_{A,1}g^{a\dagger}_\downarrow (3)+(u\overline d)^\dagger_{A,-1}g^{a\dagger}_\uparrow (3)\right]
\right\}|0\rangle \ ,
\end{eqnarray}
where $\alpha, \beta=x, y$ are the transverse indices. 
Again the color factor $T^a_{ij}/2$ is normalized to unit.
The above state obey the right transformation under $\hat Y$ because
the wave-function amplitudes are invariant when all $y$-components of 
the parton momenta change sign. 

When $|l_z|=1$, the total quark helicity must be $\lambda_{u\bar d}=0$ 
again because $\lambda_g=\pm 1$. We can write down 4
independent wave-function amplitudes, 
\begin{eqnarray}
  |\pi^+\rangle^{|l_z|=1}_{u\overline d g} &=& \int d[1]d[2]d[3] 
\frac{T^a_{ij}}{2}
\left\{
 \pn3 ^{(3)}(1,2,3)
	 \left[k_{1\perp}^+ (u\overline d)^\dagger_{A,0}g^{a\dagger}_\downarrow (3) -k_{1\perp}^-
   (u\overline d)^\dagger_{A,0}g^{a\dagger}_\uparrow (3)\right] \right. \nonumber\\
   && +\pn3 ^{(4)}(1,2,3) 
	\left[k_{2\perp}^+ (u\overline d)^\dagger_{A,0}g^{a\dagger}_\downarrow (3) -k_{2\perp}^-
   (u\overline d)^\dagger_{A,0}g^{a\dagger}_\uparrow (3)\right] \nonumber\\
   && +\pn3 ^{(5)}(1,2,3) 
	\left[k_{1\perp}^+ (u\overline d)^\dagger_{S,0}g^{a\dagger}_\downarrow (3) +k_{1\perp}^-
   (u\overline d)^\dagger_{S,0}g^{a\dagger}_\uparrow (3)\right] \nonumber\\
   && \left.+\pn3 ^{(6)}(1,2,3) 
	\left[k_{2\perp}^+ (u\overline d)^\dagger_{S,0}g^{a\dagger}_\downarrow (3) +k_{2\perp}^-
   (u\overline d)^\dagger_{S,0}g^{a\dagger}_\uparrow (3)\right] 
\right\}
 |0\rangle \ .
\end{eqnarray}

Finally, when $|l_z|=2$, the total quark helicity $\lambda_{u\bar d}=\pm 1$, 
and $\lambda_g=\pm 1$. We have 3 amplitudes
\begin{eqnarray}
  |\pi^+\rangle^{|l_z|=2}_{u\overline d g} &=& \int d[1]d[2]d[3] 
\frac{T^a_{ij}}{2}
\left\{
 \pn3 ^{(7)} (1,2,3)
   \left[k_{1\perp}^+k_{1\perp}^+ (u\overline d)^\dagger_{A,-1}g^{a\dagger}_\downarrow (3)- k_{1\perp}^-k_{1\perp}^- 
(u\overline d)^\dagger_{A,1}g^{a\dagger}_\uparrow (3)\right]\right.
   \nonumber\\
  && + \pn3 ^{(8)} (1,2,3)
   \left[k_{1\perp}^+k_{2\perp}^+ (u\overline d)^\dagger_{A,-1}g^{a\dagger}_\downarrow (3)- k_{1\perp}^-k_{2\perp}^- 
(u\overline d)^\dagger_{A,1}g^{a\dagger}_\uparrow (3)\right]
  \nonumber\\
  && \left.+\pn3 ^{(9)}(1,2,3)
   \left[k_{2\perp}^+k_{2\perp}^+ (u\overline d)^\dagger_{A,-1}g^{a\dagger}_\downarrow (3)- k_{2\perp}^-k_{2\perp}^- 
(u\overline d)^\dagger_{A,1}g^{a\dagger}_\uparrow (3)\right]\right\}
|0\rangle \ .
\end{eqnarray}
Summing up, we have a total of 9 independent light-cone amplitudes
for the pion component with 3 partons. 

\subsection{The $u\bar d gg$ Component}

For this component, the helicity of the two-gluons can be 
$\lambda_{gg}=0,~\pm 2$, and that for quarks 
$\lambda_{u\bar d}=0,~\pm 1$, and so $|l_z|= 0, 1, 2,$ or 3.
To make the $\hat Y$ transformation simple, 
we combine the two gluons in the similar way as we did 
for $u\bar d$ in the last subsection,
\begin{eqnarray}
(gg)^\dagger_{S,0}&=& g^\dagger_{\uparrow a}(3) g^\dagger_{\downarrow b}(4) 
+ g^\dagger_{\downarrow a}(3)  g^\dagger_{\uparrow b}(4) \ ,
\nonumber\\
(gg)^\dagger_{A,0}&=& g^\dagger_{\uparrow a}(3) g^\dagger_{\downarrow b}(4) 
- g^\dagger_{\downarrow a}(3) g^\dagger_{\uparrow b}(4) \ ,
\nonumber\\
(gg)^\dagger_{S,2}&=& g^\dagger_{\uparrow a}(3) g^\dagger_{\uparrow b}(4) 
\ ,\nonumber\\
(gg)^\dagger_{S,-2}&=& g^\dagger_{\downarrow a}(3) g^\dagger_{\downarrow b}(4) 
\ ,
\end{eqnarray}
where the subscripts $A$ and $S$ indicate that there is a 
factor $-1$ and $1$, respectively, under the $\hat Y$ transformation.

There are three different ways to couple the color indices
of the two quarks and two gluons to form color-singlets. If the color
indices for the two quarks are $i$ and $j$ and those for two gluons 
are $a$ and $b$, we have the singlet combinations:
$f_{abc}T^c_{ij}$, $d_{abc}T^c_{ij}$, and $\delta_{ab}\delta_{ij}$.
The last two are symmetric in the color indices of the two gluons, 
while the first one is antisymmetric. In the following, we only present 
the results for the color coupling 
$\delta_{ab}\delta_{ij}$ (the quark pair and two gluons are both
color-singlet), and those for the other two couplings can be obtained similarly.

To maximally utilize Bose symmetry between the two gluons, 
we will eliminate the momentum of the up quark (labeled by 1 below) 
in favor of the momenta of anti-down quark and the two gluons.

For $l_z=0$, the only possible parton helicity combination 
is $\lambda_{gg}=0$ and $\lambda_{u\bar d}=0$. In this case, we have 
6 independent light-cone amplitudes following Eq. (\ref{gs}),
\begin{eqnarray}
  |\pi^+\rangle^{l_z=0}_{u\overline d gg} &=& \int d[1]d[2]d[3]d[4] 
\frac{\delta_{ij}\delta^{ab}}{\sqrt{24}}
\left \{
  \p4n ^{(1)}(1,2,3,4) 
(u\overline d)^\dagger_{A,0}(gg)^\dagger_{S,0} \right.
  \nonumber \\
  && +\p4n ^{(2)}(1,2,3,4) 
(u\overline d)^\dagger_{S,0}(gg)^\dagger_{A,0} 
  \nonumber \\
  && +  i\epsilon^{\alpha\beta}k_{2\alpha}k_{3\beta}\p4n ^{(3)}(1,2,3,4)
	(u\overline d)^\dagger_{S,0}(gg)^\dagger_{S,0} 
  \nonumber \\
  && + i\epsilon^{\alpha\beta}k_{3\alpha}k_{4\beta}\p4n ^{(4)}(1,2,3,4) 
	(u\overline d)^\dagger_{S,0}(gg)^\dagger_{S,0} 
  \nonumber \\
  && + i\epsilon^{\alpha\beta}k_{2\alpha}k_{3\beta}\p4n ^{(5)}(1,2,3,4) 
	(u\overline d)^\dagger_{A,0}(gg)^\dagger_{A,0} 
  \nonumber \\
  && \left.+ i\epsilon^{\alpha\beta}k_{3\alpha}k_{4\beta}\p4n ^{(6)}(1,2,3,4) 
	(u\overline d)^\dagger_{A,0}(gg)^\dagger_{A,0} \right \}
 |0\rangle \ ,
\end{eqnarray}
where we have used the symmetry between two gluons ($3\leftrightarrow 4$) 
to reduce the number of independent amplitudes.
For example, $i\epsilon^{\alpha\beta}k_{2\alpha}k_{4\beta}$ 
can be obtained from $i\epsilon^{\alpha\beta}k_{2\alpha}k_{3\beta}$
by 3 and 4 exchange, and so the former is not independent.
This property is general for all of the light-cone amplitudes
of $u\overline d gg$ component, and will be used throughout the
following classification. Because of (anti)symmetric properties for the two
gluons, the above amplitudes have the following symmetry: 
$\p4n ^{(1,6)}(1,2,3,4)=\p4n ^{(1,6)}(1,2,4,3)$ and 
$\p4n ^{(2,4)}(1,2,3,4)=-\p4n ^{(2,4)}(1,2,4,3)$.

For $|l_z|=1$, the parton helicity has two possible combinations:
either $\lambda_{gg}=0$ and $\lambda_{u\bar d}=\mp 1$, or
$\lambda_{gg}=\mp 2$ and $\lambda_{u\bar d}=\pm 1$.
For the first case, we have 8 independent amplitudes,
\begin{eqnarray}
  |\pi^+\rangle^{|l_z|=1}_{u\overline d gg} &=& \int d[1]d[2]d[3]d[4] 
\frac{\delta^{ij}\delta_{ab}}{\sqrt{24}}\left\{\right.\nonumber\\
&&~~\p4n ^{(7)} (1,2,3,4) 
\left[k_{2\perp}^+ (u\overline d)^\dagger_{A,-1}(gg)^\dagger_{S,0} +k_{2\perp}^{-}
     (u\overline d)^\dagger_{A,1}(gg)^\dagger_{S,0}\right]
   \nonumber\\
&& + \p4n ^{(8)}(1,2,3,4) 
\left[k_{3\perp}^+ (u\overline d)^\dagger_{A,-1}(gg)^\dagger_{S,0} +k_{3\perp}^{-}
     (u\overline d)^\dagger_{A,1}(gg)^\dagger_{S,0}\right]
   \nonumber \\
      && + \p4n ^{(9)} (1,2,3,4) 
\left[k_{2\perp}^+ (u\overline d)^\dagger_{A,-1}(gg)^\dagger_{A,0} -k_{2\perp}^{-}
     (u\overline d)^\dagger_{A,1}(gg)^\dagger_{A,0}\right]
   \nonumber\\
&& + \p4n ^{(10)}(1,2,3,4) 
\left[k_{3\perp}^+ (u\overline d)^\dagger_{A,-1}(gg)^\dagger_{A,0} -k_{3\perp}^{-}
     (u\overline d)^\dagger_{A,1}(gg)^\dagger_{A,0}\right]
    \\
   && + i\epsilon^{\alpha\beta}k_{3\alpha}k_{4\beta}
	\p4n ^{(11)} (1,2,3,4) 
	\left[k_{2\perp}^+ (u\overline d)^\dagger_{A,-1}(gg)^\dagger_{S,0} -k_{2\perp}^{-}
     (u\overline d)^\dagger_{A,1}(gg)^\dagger_{S,0}\right]
   \nonumber\\
&&  + i\epsilon^{\alpha\beta}k_{2\alpha}k_{4\beta}
	\p4n ^{(12)}(1,2,3,4) 
	\left[k_{3\perp}^+ (u\overline d)^\dagger_{A,-1}(gg)^\dagger_{S,0} -k_{3\perp}^{-}
     (u\overline d)^\dagger_{A,1}(gg)^\dagger_{S,0}\right]
   \nonumber \\
   && + i\epsilon^{\alpha\beta}k_{3\alpha}k_{4\beta}
	\p4n ^{(13)} (1,2,3,4) 
	\left[k_{2\perp}^+ (u\overline d)^\dagger_{A,-1}(gg)^\dagger_{A,0} +k_{2\perp}^{-}
     (u\overline d)^\dagger_{A,1}(gg)^\dagger_{A,0}\right]
   \nonumber\\
&& \left. + i\epsilon^{\alpha\beta}k_{2\alpha}k_{4\beta}
	\p4n ^{(14)}(1,2,3,4) 
	\left[k_{3\perp}^+ (u\overline d)^\dagger_{A,-1}(gg)^\dagger_{A,0} +k_{3\perp}^{-}
     (u\overline d)^\dagger_{A,1}(gg)^\dagger_{A,0}\right]
\right\}
 |0\rangle \ , \nonumber 
\end{eqnarray}
where the $3\leftrightarrow 4$ symmetry again plays an important role to 
reduce the number of independent amplitudes.
For the second case, $\lambda_{gg}=\mp 2$ and $\lambda_{u\bar d}=\pm 1$, 
we find 4 independent amplitudes:
\begin{eqnarray}
  |\pi^+\rangle^{|l_z|=1}_{u\overline d gg} &=& \int d[1]d[2]d[3]d[4] 
\frac{\delta^{ij}\delta_{ab}}{\sqrt{24}}\left\{\right.\nonumber\\
&&~~\p4n ^{(15)} (1,2,3,4) 
	\left[k_{2\perp}^+ (u\overline d)^\dagger_{A,1}(gg)^\dagger_{S,-2} +k_{2\perp}^{-}
     		(u\overline d)^\dagger_{A,-1}(gg)^\dagger_{S,2}\right]
   \nonumber\\
&& + \p4n ^{(16)}(1,2,3,4) 
	\left[k_{3\perp}^+ (u\overline d)^\dagger_{A,1}(gg)^\dagger_{S,-2} +k_{3\perp}^{-}
     		(u\overline d)^\dagger_{A,-1}(gg)^\dagger_{S,2}\right]
      \\
   && + i\epsilon^{\alpha\beta}k_{3\alpha}k_{4\beta}
	\p4n ^{(17)} (1,2,3,4) 
	\left[k_{2\perp}^+ (u\overline d)^\dagger_{A,1}(gg)^\dagger_{S,-2} -k_{2\perp}^{-}
     		(u\overline d)^\dagger_{A,-1}(gg)^\dagger_{S,2}\right]
   \nonumber\\
   && \left.+ i\epsilon^{\alpha\beta}k_{3\alpha}k_{4\beta}
	\p4n ^{(18)} (1,2,3,4) 
	\left[k_{3\perp}^+ (u\overline d)^\dagger_{A,1}(gg)^\dagger_{S,-2} -k_{3\perp}^{-}
     		(u\overline d)^\dagger_{A,-1}(gg)^\dagger_{S,2}\right]
\right\}
 |0\rangle \ .\nonumber
\end{eqnarray}
The Bose symmetry implies the following constraints:
$\p4n ^{(7,13,15)} (1,2,3,4)=\p4n ^{(7,13,15)} (1,2,4,3)$,
and $\p4n ^{(8,11,17)} (1,2,3,4)=-\p4n ^{(8,11,17)} (1,2,4,3)$.

For $|l_z|=2$, the parton helicity must be 
$\lambda_{gg}=\mp 2$ and $\lambda_{u\bar d}=0$. We find 12 independent amplitudes
\begin{eqnarray}
  |\pi^+\rangle^{|l_z|=2}_{u\overline d gg} &=& \int d[1]d[2]d[3]d[4] 
\frac{\delta^{ij}\delta_{ab}}{\sqrt{24}}\left\{\right.\nonumber\\
&&~~\p4n ^{(19)}(1,2,3,4) 
	\left[k_{2\perp}^+k_{2\perp}^+(u\overline d)^\dagger_{A,0}(gg)^\dagger_{S,-2}
  	+ k_{2\perp}^-k_{2\perp}^-(u\overline d)^\dagger_{A,0}(gg)^\dagger_{S,2}\right]
	  \nonumber\\
  && + \p4n ^{(20)}(1,2,3,4) 
\left[k_{2\perp}^+k_{3\perp}^+(u\overline d)^\dagger_{A,0}(gg)^\dagger_{S,-2}
  + k_{2\perp}^-k_{3\perp}^-(u\overline d)^\dagger_{A,0}(gg)^\dagger_{S,2}\right]
\nonumber\\
 && + \p4n ^{(21)}(1,2,3,4) 
\left[k_{3\perp}^+k_{3\perp}^+(u\overline d)^\dagger_{A,0}(gg)^\dagger_{S,-2}
  + k_{3\perp}^-k_{3\perp}^-(u\overline d)^\dagger_{A,0}(gg)^\dagger_{S,2}\right]
     \nonumber\\
 && + \p4n ^{(22)}(1,2,3,4) 
\left[k_{3\perp}^+k_{4\perp}^+(u\overline d)^\dagger_{A,0}(gg)^\dagger_{S,-2}
  + k_{3\perp}^-k_{4\perp}^-(u\overline d)^\dagger_{A,0}(gg)^\dagger_{S,2}\right]
\nonumber
\end{eqnarray}
\begin{eqnarray}
&&+ i\epsilon^{\alpha\beta}k_{3\alpha}k_{4\beta}
\p4n ^{(23)}(1,2,3,4) 
\left[k_{2\perp}^+k_{2\perp}^+(u\overline d)^\dagger_{A,0}(gg)^\dagger_{S,-2}
  - k_{2\perp}^-k_{2\perp}^-(u\overline d)^\dagger_{A,0}(gg)^\dagger_{S,2}\right]
  \nonumber\\
&&+ i\epsilon^{\alpha\beta}k_{2\alpha}k_{4\beta}
\p4n ^{(24)}(1,2,3,4) 
\left[k_{3\perp}^+k_{3\perp}^+(u\overline d)^\dagger_{A,0}(gg)^\dagger_{S,-2}
  - k_{3\perp}^-k_{3\perp}^-(u\overline d)^\dagger_{A,0}(gg)^\dagger_{S,2}\right]
  \nonumber\\
&& + \p4n ^{(25)}(1,2,3,4) 
\left[k_{2\perp}^+k_{2\perp}^+(u\overline d)^\dagger_{S,0}(gg)^\dagger_{S,-2}
  - k_{2\perp}^-k_{2\perp}^-(u\overline d)^\dagger_{S,0}(gg)^\dagger_{S,2}\right]
  \\
&& + \p4n ^{(26)}(1,2,3,4) 
\left[k_{2\perp}^+k_{3\perp}^+(u\overline d)^\dagger_{S,0}(gg)^\dagger_{S,-2}
  - k_{2\perp}^-k_{3\perp}^-(u\overline d)^\dagger_{S,0}(gg)^\dagger_{S,2}\right]
  \nonumber\\
&& + \p4n ^{(27)}(1,2,3,4) 
\left[k_{3\perp}^+k_{3\perp}^+(u\overline d)^\dagger_{S,0}(gg)^\dagger_{S,-2}
  - k_{3\perp}^-k_{3\perp}^-(u\overline d)^\dagger_{S,0}(gg)^\dagger_{S,2}\right]
  \nonumber\\
&& + \p4n ^{(28)}(1,2,3,4) 
\left[k_{3\perp}^+k_{4\perp}^+(u\overline d)^\dagger_{S,0}(gg)^\dagger_{S,-2}
  - k_{3\perp}^-k_{4\perp}^-(u\overline d)^\dagger_{S,0}(gg)^\dagger_{S,2}\right]
  \nonumber\\
&& + i\epsilon^{\alpha\beta}k_{3\alpha}k_{4\beta}
\p4n ^{(29)}(1,2,3,4) 
\left[k_{2\perp}^+k_{2\perp}^+(u\overline d)^\dagger_{S,0}(gg)^\dagger_{S,-2}
  + k_{2\perp}^-k_{2\perp}^-(u\overline d)^\dagger_{S,0}(gg)^\dagger_{S,2}\right]
  \nonumber\\
&&\left. + i\epsilon^{\alpha\beta}k_{2\alpha}k_{4\beta}
\p4n ^{(30)}(1,2,3,4) 
\left[k_{3\perp}^+k_{3\perp}^+(u\overline d)^\dagger_{S,0}(gg)^\dagger_{S,-2}
  + k_{3\perp}^-k_{3\perp}^-(u\overline d)^\dagger_{S,0}(gg)^\dagger_{S,2}\right]
\right\}
 |0\rangle \ .
\nonumber
\end{eqnarray}
The Bose symmetry yields the following constraints:
$\p4n ^{(19,22,25,28)}(1,2,3,4)=\p4n ^{(19,22,25,28)}(1,2,4,3)$
and $\p4n ^{(23,29)}(1,2,3,4)=-\p4n ^{(23,29)}(1,2,4,3)$.

For $|l_z|=3$, the parton helicity must be $\lambda_{gg}=\pm 2$ and 
$\lambda_{u\bar d}=\pm 1$. We find 8 independent light-cone amplitudes
\begin{eqnarray}
  |\pi^+\rangle^{|l_z|=3}_{u\overline d gg} &=& \int d[1]d[2]d[3]d[4] 
\frac{\delta^{ij}\delta_{ab}}{\sqrt{24}}\left\{\right.\nonumber\\
&&~~\p4n ^{(31)}(1,2,3,4)\left[(k_{2\perp}^+)^3 (u\overline d)^\dagger_{A,-1}(gg)^\dagger_{S,-2}
                   +(k_{2\perp}^-)^3 (u\overline d)^\dagger_{A,1}(gg)^\dagger_{S,2}\right]
  \nonumber\\
  && + \p4n ^{(32)}(1,2,3,4)\left[(k_{3\perp}^+)^3 (u\overline d)^\dagger_{A,-1}(gg)^\dagger_{S,-2}
                   +(k_{3\perp}^-)^3 (u\overline d)^\dagger_{A,1}(gg)^\dagger_{S,2}\right]
  \nonumber\\
  &&  + \p4n ^{(33)}(1,2,3,4)\left[(k_{2\perp}^+)^2k_{3\perp}^+ (u\overline d)^\dagger_{A,-1}(gg)^\dagger_{S,-2}
                   +(k_{2\perp}^-)^2k_{3\perp}^- (u\overline d)^\dagger_{A,1}(gg)^\dagger_{S,2}\right]
  \nonumber\\
  && + \p4n ^{(34)}(1,2,3,4)\left[(k_{3\perp}^+)^2k_{2\perp}^+ (u\overline d)^\dagger_{A,-1}(gg)^\dagger_{S,-2}
                   +(k_{3\perp}^-)^2k_{2\perp}^- (u\overline d)^\dagger_{A,1}(gg)^\dagger_{S,2}\right]
	\\
  && + \p4n ^{(35)}(1,2,3,4)\left[(k_{3\perp}^+)^2k_{4\perp}^+ (u\overline d)^\dagger_{A,-1}(gg)^\dagger_{S,-2}
                   +(k_{3\perp}^-)^2k_{4\perp}^- (u\overline d)^\dagger_{A,1}(gg)^\dagger_{S,2}\right]
                   \nonumber\\
    && + \p4n ^{(36)}(1,2,3,4)\left[k_{2\perp}^+k_{3\perp}^+k_{4\perp}^+ (u\overline d)^\dagger_{A,-1}(gg)^\dagger_{S,-2}
                   +k_{2\perp}^-k_{3\perp}^-k_{4\perp}^- (u\overline d)^\dagger_{A,1}(gg)^\dagger_{S,2}\right]
   \nonumber\\
  && +i\epsilon^{\alpha\beta}k_{3\alpha}k_{4\beta}
\p4n ^{(37)}(1,2,3,4)\left[(k_{2\perp}^+)^3 (u\overline d)^\dagger_{A,-1}(gg)^\dagger_{S,-2}
                   -(k_{2\perp}^-)^3 (u\overline d)^\dagger_{A,1}(gg)^\dagger_{S,2}\right]
  \nonumber\\
  &&\left. +i\epsilon^{\alpha\beta}k_{2\alpha}k_{4\beta}
\p4n ^{(38)}(1,2,3,4)\left[(k_{3\perp}^+)^3 (u\overline d)^\dagger_{A,-1}(gg)^\dagger_{S,-2}
                   -(k_{3\perp}^-)^3 (u\overline d)^\dagger_{A,1}(gg)^\dagger_{S,2}\right]
	\right\}
 |0\rangle \ .
\nonumber
\end{eqnarray}
The Bose symmetry implies the symmetry relations:
$\p4n ^{(31,36)}(1,2,3,4)=\p4n ^{(31,36)}(1,2,4,3)$
and $\p4n ^{(37)}(1,2,3,4)=-\p4n ^{(37)}(1,2,4,3)$.

Similarly, one can obtain the amplitudes when the quark pair and two
gluons are in color-octet states. If 
the two gluons are symmetric in color, we have
$\p4n ^{(i)}$ with $i=39,\cdots, 76$, defined in the same
way as the above equations except the color factor is replaced
by $\sqrt{\frac{3}{20}}d_{abc}T^c_{ij}$. 
When the two gluons are antisymmetric in color, we obtain
$\p4n ^{(i)}$ with $i=77,\cdots, 114$, again defined in the same
way, except with the color factor $\sqrt{\frac{1}{12}}f_{abc}T^c_{ij}$.
Note that there are sign changes for the symmetry relations 
derived from Bose symmetry.

Therefore, we have a total of $38\times 3=114$ independent amplitudes 
for the Fock component $u\overline d gg$ in $\pi^+$.

\subsection{The $u\bar d q\bar q$ Component}

We first consider the up and down sea-quark flavors. 
In this case, the following two flavor 
structures have total isospin $I=1$: 
\begin{eqnarray}
&&u\overline d (u\overline u + d\overline d) \ ,\nonumber\\
&&(u\overline u-d\overline d)u\overline d-u\overline d(u\overline u-d\overline d) \ .
\end{eqnarray}
The first structure arises from the first quark pair coupled to
isospin 1 and the second pair coupled to isospin 0. The second
structure comes from both pairs coupled to isospin 1.  
However, after some rearrangements of 
the particle labels, the second structure can be reduced to the 
first one, and hence is not independent. 
Therefore, we consider only the first isospin structure 
with all possible color and spin combinations.

To simplify the $\hat Y$ transformation, we introduce
the following combinations for the sea quark pair:
\begin{eqnarray}
(q\overline q)^\dagger_{S,0}&=& u^\dagger_{\uparrow k}(3)\overline u^\dagger_{\downarrow l}(4) 
+ u^\dagger_{\downarrow k}(3) \overline u^\dagger_{\uparrow l}(4) +
d^\dagger_{\uparrow k}(3)\overline d^\dagger_{\downarrow l}(4) 
+ d^\dagger_{\downarrow k}(3) \overline d^\dagger_{\uparrow l}(4)\ ,
\nonumber\\
(q\overline q)^\dagger_{A,0}&=& u^\dagger_{\uparrow k}(3)\overline u^\dagger_{\downarrow l}(4) 
- u^\dagger_{\downarrow k}(3) \overline u^\dagger_{\uparrow l}(4) +
d^\dagger_{\uparrow k}(3)\overline d^\dagger_{\downarrow l}(4) 
- d^\dagger_{\downarrow k}(3) \overline d^\dagger_{\uparrow l}(4)\ ,
\nonumber\\
(q\overline q)^\dagger_{A,1}&=& u^\dagger_{\uparrow k}(3)\overline u^\dagger_{\uparrow l}(4) 
 +d^\dagger_{\uparrow k}(3)\overline d^\dagger_{\uparrow l}(4) \ ,
\nonumber\\
(q\overline q)^\dagger_{A,-1}&=& u^\dagger_{\downarrow k}(3)\overline u^\dagger_{\downarrow l}(4) 
+d^\dagger_{\downarrow k}(3)\overline d^\dagger_{\downarrow l}(4) 
\ .
\end{eqnarray}
We use them as basic building blocks in the Fock expansion.

We can form two color-singlet structures from  
the four color indices $i$,$j$,$k$, and $l$: $\delta_{ij}\delta_{kl}$
and $\delta_{il}\delta_{jk}$, where we have implicitly assumed that
the first and third are quarks' and second and fourth are antiquarks'. 
The first structure corresponds to the state in which the two quark-pairs 
are both coupled to color-singlet, while the second corresponds to the state 
in which the two quark-pairs are in color-octet. The wave-function
amplitudes for both color combinations are similar.

The quark helicity has combinations $\lambda_{u\bar d}=0,\ \pm 1$, 
and $\lambda_{u\bar u+d\bar d}=0,\ \pm 1$. Therefore we can 
have three different orbital angular momentum projection $|l_z|=0,1,2$.
 
For $l_z=0$, the quark helicity has the combination 
$\lambda_{u\bar d}=0$ and $\lambda_{u\bar u+d\bar d}=0$,
or $\lambda_{u\bar d}=\pm 1$ and $\lambda_{u\bar u+d\bar d}=\mp 1$. Together, 
we find 12 independent amplitudes:
\begin{eqnarray}
  |\pi^+\rangle^{l_z=0}_{u\bar d q\bar q} &=& \int d[1]d[2]d[3]d[4] 
\frac{\delta_{ij}\delta_{kl}}{3}\left\{ \right. \nonumber\\
&&~~ \pu4 ^{(1)} (1,2,3,4) (u\overline d)^\dagger_{A,0}(q\overline q)^\dagger_{S,0}+
  \pu4 ^{(2)} (1,2,3,4) (u\overline d)^\dagger_{S,0}(q\overline q)^\dagger_{A,0}
  \nonumber\\
  && +i\epsilon^{\alpha\beta}k_{1\alpha}k_{2\beta}
	\pu4 ^{(3)} (1,2,3,4) (u\overline d)^\dagger_{S,0}(q\overline q)^\dagger_{S,0}
  \nonumber\\
  && +i\epsilon^{\alpha\beta}k_{1\alpha}k_{3\beta}
	\pu4 ^{(4)} (1,2,3,4) (u\overline d)^\dagger_{S,0}(q\overline q)^\dagger_{S,0}
  \nonumber\\
  && +i\epsilon^{\alpha\beta}k_{2\alpha}k_{3\beta}
	\pu4 ^{(5)} (1,2,3,4) (u\overline d)^\dagger_{S,0}(q\overline q)^\dagger_{S,0}
  \nonumber\\
  && +i\epsilon^{\alpha\beta}k_{1\alpha}k_{2\beta}
	\pu4 ^{(6)} (1,2,3,4) (u\overline d)^\dagger_{A,0}(q\overline q)^\dagger_{A,0}
  \nonumber
\end{eqnarray}
\begin{eqnarray}
  && +i\epsilon^{\alpha\beta}k_{1\alpha}k_{3\beta}
	\pu4 ^{(7)} (1,2,3,4) (u\overline d)^\dagger_{A,0}(q\overline q)^\dagger_{A,0}
 \nonumber\\
  && +i\epsilon^{\alpha\beta}k_{2\alpha}k_{3\beta}
	\pu4 ^{(8)} (1,2,3,4) (u\overline d)^\dagger_{A,0}(q\overline q)^\dagger_{A,0}
  \nonumber\\
  && +\pu4 ^{(9)} (1,2,3,4) \left[(u\overline d)^\dagger_{A,1}(q\overline q)^\dagger_{A,-1}
      - (u\overline d)^\dagger_{A,-1}(q\overline q)^\dagger_{A,1}\right]
  \nonumber\\
  && +i\epsilon^{\alpha\beta}k_{1\alpha}k_{2\beta}
	\pu4 ^{(10)} (1,2,3,4) \left[(u\overline d)^\dagger_{A,1}(q\overline q)^\dagger_{A,-1}
      + (u\overline d)^\dagger_{A,-1}(q\overline q)^\dagger_{A,1}\right]
  \nonumber\\
 && +i\epsilon^{\alpha\beta}k_{1\alpha}k_{3\beta}
	\pu4 ^{(11)} (1,2,3,4) \left[(u\overline d)^\dagger_{A,1}(q\overline q)^\dagger_{A,-1}
      + (u\overline d)^\dagger_{A,-1}(q\overline q)^\dagger_{A,1}\right]
  \nonumber\\
&& \left. +i\epsilon^{\alpha\beta}k_{2\alpha}k_{3\beta}
	\pu4 ^{(12)} (1,2,3,4) \left[(u\overline d)^\dagger_{A,1}(q\overline q)^\dagger_{A,-1}
      + (u\overline d)^\dagger_{A,-1}(q\overline q)^\dagger_{A,1}\right]
\right\}
|0\rangle \ .
\end{eqnarray}
Note that $\delta_{ij}$ implicitly contracts the color indices in the $u\bar d$ pair
and $\delta_{kl}$ contracts the $q\bar q$ pair.

For $|l_z|=1$, the quark helicity can either be in the
combination $\lambda_{u\bar d}=0$ and $\lambda_{u\bar u}=-1$, or
$\lambda_{u\bar d}=-1$ and $\lambda_{u\bar u}=0$. Taking together,
we find 24 independent amplitudes:
\begin{eqnarray}
  |\pi^+\rangle^{|l_z|=1}_{u\bar d q\bar q} &=& \int d[1]d[2]d[3]d[4] 
\frac{\delta_{ij}\delta_{kl}}{3}\left\{ \right. \nonumber\\
 && ~~\pu4 ^{(13)} (1,2,3,4) \left[k_{1\perp}^+(u\overline d)^\dagger_{A,0}(q\overline q)^\dagger_{A,-1} -
                               k_{1\perp}^-(u\overline d)^\dagger_{A,0}(q\overline q)^\dagger_{A,1}\right]
 \nonumber\\
 && + \pu4 ^{(14)} (1,2,3,4) \left[k_{1\perp}^+(u\overline d)^\dagger_{S,0}(q\overline q)^\dagger_{A,-1} +
                               k_{1\perp}^-(u\overline d)^\dagger_{S,0}(q\overline q)^\dagger_{A,1}\right]
\nonumber\\
 && + \pu4 ^{(15)} (1,2,3,4) \left[k_{1\perp}^+(u\overline d)^\dagger_{A,-1}(q\overline q)^\dagger_{A,0} -
                               k_{1\perp}^-(u\overline d)^\dagger_{A,1}(q\overline q)^\dagger_{A,0}\right]
 \nonumber\\
 && + \pu4 ^{(16)} (1,2,3,4) \left[k_{1\perp}^+(u\overline d)^\dagger_{A,-1}(q\overline q)^\dagger_{S,0} +
                               k_{1\perp}^-(u\overline d)^\dagger_{A,1}(q\overline q)^\dagger_{S,0}\right]
  \nonumber\\
 && +\pu4 ^{(17)} (1,2,3,4) \left[k_{2\perp}^+(u\overline d)^\dagger_{A,0}(q\overline q)^\dagger_{A,-1} -
                               k_{2\perp}^-(u\overline d)^\dagger_{A,0}(q\overline q)^\dagger_{A,1}\right]
 \nonumber\\
 && + \pu4 ^{(18)} (1,2,3,4) \left[k_{2\perp}^+(u\overline d)^\dagger_{S,0}(q\overline q)^\dagger_{A,-1} +
                               k_{2\perp}^-(u\overline d)^\dagger_{S,0}(q\overline q)^\dagger_{A,1}\right]
\nonumber\\
 && + \pu4 ^{(19)} (1,2,3,4) \left[k_{2\perp}^+(u\overline d)^\dagger_{A,-1}(q\overline q)^\dagger_{A,0} -
                               k_{2\perp}^-(u\overline d)^\dagger_{A,1}(q\overline q)^\dagger_{A,0}\right]
 \nonumber\\
 && + \pu4 ^{(20)} (1,2,3,4) \left[k_{2\perp}^+(u\overline d)^\dagger_{A,-1}(q\overline q)^\dagger_{S,0} +
                               k_{2\perp}^-(u\overline d)^\dagger_{A,1}(q\overline q)^\dagger_{S,0}\right]
  \nonumber\\
 && +\pu4 ^{(21)} (1,2,3,4) \left[k_{3\perp}^+(u\overline d)^\dagger_{A,0}(q\overline q)^\dagger_{A,-1} -
                               k_{3\perp}^-(u\overline d)^\dagger_{A,0}(q\overline q)^\dagger_{A,1}\right]
 \nonumber\\
 && + \pu4 ^{(22)} (1,2,3,4) \left[k_{3\perp}^+(u\overline d)^\dagger_{S,0}(q\overline q)^\dagger_{A,-1} +
                               k_{3\perp}^-(u\overline d)^\dagger_{S,0}(q\overline q)^\dagger_{A,1}\right]
\nonumber\\
 && + \pu4 ^{(23)} (1,2,3,4) \left[k_{3\perp}^+(u\overline d)^\dagger_{A,-1}(q\overline q)^\dagger_{A,0} -
                               k_{3\perp}^-(u\overline d)^\dagger_{A,1}(q\overline q)^\dagger_{A,0}\right]
 \nonumber\\
 && + \pu4 ^{(24)} (1,2,3,4) \left[k_{3\perp}^+(u\overline d)^\dagger_{A,-1}(q\overline q)^\dagger_{S,0} +
                               k_{3\perp}^-(u\overline d)^\dagger_{A,1}(q\overline q)^\dagger_{S,0}\right]
  \nonumber \\
 && +i\epsilon^{\alpha\beta}k_{2\alpha}k_{3\beta}
	\pu4 ^{(25)} (1,2,3,4) \left[k_{1\perp}^+(u\overline d)^\dagger_{A,0}(q\overline q)^\dagger_{A,-1} +
                               k_{1\perp}^-(u\overline d)^\dagger_{A,0}(q\overline q)^\dagger_{A,1}\right]
 \nonumber \\
 && + i\epsilon^{\alpha\beta}k_{2\alpha}k_{3\beta}
	\pu4 ^{(26)} (1,2,3,4) \left[k_{1\perp}^+(u\overline d)^\dagger_{S,0}(q\overline q)^\dagger_{A,-1} -
                               k_{1\perp}^-(u\overline d)^\dagger_{S,0}(q\overline q)^\dagger_{A,1}\right]
\nonumber\\
 && + i\epsilon^{\alpha\beta}k_{2\alpha}k_{3\beta}
	\pu4 ^{(27)} (1,2,3,4) \left[k_{1\perp}^+(u\overline d)^\dagger_{A,-1}(q\overline q)^\dagger_{A,0} +
                               k_{1\perp}^-(u\overline d)^\dagger_{A,1}(q\overline q)^\dagger_{A,0}\right]
 \nonumber\\
 && + i\epsilon^{\alpha\beta}k_{2\alpha}k_{3\beta}
	\pu4 ^{(28)} (1,2,3,4) \left[k_{1\perp}^+(u\overline d)^\dagger_{A,-1}(q\overline q)^\dagger_{S,0} -
                               k_{1\perp}^-(u\overline d)^\dagger_{A,1}(q\overline q)^\dagger_{S,0}\right]
  \nonumber
\end{eqnarray}
\begin{eqnarray}
 && +i\epsilon^{\alpha\beta}k_{1\alpha}k_{3\beta}
	\pu4 ^{(29)} (1,2,3,4) \left[k_{2\perp}^+(u\overline d)^\dagger_{A,0}(q\overline q)^\dagger_{A,-1} +
                               k_{2\perp}^-(u\overline d)^\dagger_{A,0}(q\overline q)^\dagger_{A,1}\right]
 \nonumber\\
 && + i\epsilon^{\alpha\beta}k_{1\alpha}k_{3\beta}
	\pu4 ^{(30)} (1,2,3,4) \left[k_{2\perp}^+(u\overline d)^\dagger_{S,0}(q\overline q)^\dagger_{A,-1} -
                               k_{2\perp}^-(u\overline d)^\dagger_{S,0}(q\overline q)^\dagger_{A,1}\right]
\nonumber\\
 && + i\epsilon^{\alpha\beta}k_{1\alpha}k_{3\beta}
	\pu4 ^{(31)} (1,2,3,4) \left[k_{2\perp}^+(u\overline d)^\dagger_{A,-1}(q\overline q)^\dagger_{A,0} +
                               k_{2\perp}^-(u\overline d)^\dagger_{A,1}(q\overline q)^\dagger_{A,0}\right]
 \nonumber\\
 && + i\epsilon^{\alpha\beta}k_{1\alpha}k_{3\beta}
	\pu4 ^{(32)} (1,2,3,4) \left[k_{2\perp}^+(u\overline d)^\dagger_{A,-1}(q\overline q)^\dagger_{S,0} -
                               k_{2\perp}^-(u\overline d)^\dagger_{A,1}(q\overline q)^\dagger_{S,0}\right]
  \nonumber\\
 && +i\epsilon^{\alpha\beta}k_{1\alpha}k_{2\beta}
	\pu4 ^{(33)} (1,2,3,4) \left[k_{3\perp}^+(u\overline d)^\dagger_{A,0}(q\overline q)^\dagger_{A,-1} +
                               k_{3\perp}^-(u\overline d)^\dagger_{A,0}(q\overline q)^\dagger_{A,1}\right]
 \nonumber\\
 && + i\epsilon^{\alpha\beta}k_{1\alpha}k_{2\beta}
	\pu4 ^{(34)} (1,2,3,4) \left[k_{3\perp}^+(u\overline d)^\dagger_{S,0}(q\overline q)^\dagger_{A,-1} -
                               k_{3\perp}^-(u\overline d)^\dagger_{S,0}(q\overline q)^\dagger_{A,1}\right]
\nonumber\\
 && + i\epsilon^{\alpha\beta}k_{1\alpha}k_{2\beta}
	\pu4 ^{(35)} (1,2,3,4) \left[k_{3\perp}^+(u\overline d)^\dagger_{A,-1}(q\overline q)^\dagger_{A,0} +
                               k_{3\perp}^-(u\overline d)^\dagger_{A,1}(q\overline q)^\dagger_{A,0}\right]
 \nonumber\\
 && \left. + i\epsilon^{\alpha\beta}k_{1\alpha}k_{2\beta}
	\pu4 ^{(36)} (1,2,3,4) \left[k_{3\perp}^+(u\overline d)^\dagger_{A,-1}(q\overline q)^\dagger_{S,0} -
                               k_{3\perp}^-(u\overline d)^\dagger_{A,1}(q\overline q)^\dagger_{S,0}\right]
\right\}
|0\rangle \ .
\end{eqnarray}

For $|l_z|=2$, the quark helicity must be $\lambda_{u\bar d}=\pm 1$ and $\lambda_{u\bar u}=\pm 1$.
We have the following 9 independent amplitudes,
\begin{eqnarray}
  |\pi^+\rangle^{|l_z|=2}_{u\bar d q\bar q} &=& \int d[1]d[2]d[3]d[4] 
\frac{\delta_{ij}\delta_{kl}}{3}\left\{ \right. \nonumber\\
 && ~~ \pu4 ^{(37)} (1,2,3,4) \left[k_{1\perp}^+k_{1\perp}^+(u\overline d)^\dagger_{A,-1}(q\overline q)^\dagger_{A,-1} -
                               k_{1\perp}^-k_{1\perp}^-(u\overline d)^\dagger_{A,1}(q\overline q)^\dagger_{A,1}\right]
 \nonumber\\
   &&  + \pu4 ^{(38)} (1,2,3,4) \left[k_{2\perp}^+k_{2\perp}^+(u\overline d)^\dagger_{A,-1}(q\overline q)^\dagger_{A,-1} -
                               k_{2\perp}^-k_{2\perp}^-(u\overline d)^\dagger_{A,1}(q\overline q)^\dagger_{A,1}\right]
 \nonumber\\
  && +  \pu4 ^{(39)} (1,2,3,4) \left[k_{3\perp}^+k_{3\perp}^+(u\overline d)^\dagger_{A,-1}(q\overline q)^\dagger_{A,-1} -
                               k_{3\perp}^-k_{3\perp}^-(u\overline d)^\dagger_{A,1}(q\overline q)^\dagger_{A,1}\right]
 \nonumber\\
 && + \pu4 ^{(40)} (1,2,3,4) \left[k_{1\perp}^+k_{2\perp}^+(u\overline d)^\dagger_{A,-1}(q\overline q)^\dagger_{A,-1} -
                               k_{1\perp}^-k_{2\perp}^-(u\overline d)^\dagger_{A,1}(q\overline q)^\dagger_{A,1}\right]
 \nonumber\\
   &&  + \pu4 ^{(41)} (1,2,3,4) \left[k_{1\perp}^+k_{3\perp}^+(u\overline d)^\dagger_{A,-1}(q\overline q)^\dagger_{A,-1} -
                               k_{1\perp}^-k_{3\perp}^-(u\overline d)^\dagger_{A,1}(q\overline q)^\dagger_{A,1}\right]
\\
  && +  \pu4 ^{(42)} (1,2,3,4) \left[k_{2\perp}^+k_{3\perp}^+(u\overline d)^\dagger_{A,-1}(q\overline q)^\dagger_{A,-1} -
                               k_{2\perp}^-k_{3\perp}^-(u\overline d)^\dagger_{A,1}(q\overline q)^\dagger_{A,1}\right]
 \nonumber\\
 && + i\epsilon^{\alpha\beta}k_{2\alpha}k_{3\beta}
	\pu4 ^{(43)} (1,2,3,4) \left[k_{1\perp}^+k_{1\perp}^+(u\overline d)^\dagger_{A,-1}(q\overline q)^\dagger_{A,-1} +
                               k_{1\perp}^-k_{1\perp}^-(u\overline d)^\dagger_{A,1}(q\overline q)^\dagger_{A,1}\right]
 \nonumber\\
   &&  + i\epsilon^{\alpha\beta}k_{1\alpha}k_{3\beta}
	\pu4 ^{(44)} (1,2,3,4) \left[k_{2\perp}^+k_{2\perp}^+(u\overline d)^\dagger_{A,-1}(q\overline q)^\dagger_{A,-1} +
                               k_{2\perp}^-k_{2\perp}^-(u\overline d)^\dagger_{A,1}(q\overline q)^\dagger_{A,1}\right]
 \nonumber\\
  && \left. +  i\epsilon^{\alpha\beta}k_{1\alpha}k_{2\beta}
	\pu4 ^{(45)} (1,2,3,4) \left[k_{3\perp}^+k_{3\perp}^+(u\overline d)^\dagger_{A,-1}(q\overline q)^\dagger_{A,-1} +
                               k_{3\perp}^-k_{3\perp}^-(u\overline d)^\dagger_{A,1}(q\overline q)^\dagger_{A,1}\right]
	\right\}|0\rangle \ . 
 \nonumber
\end{eqnarray}

In summary, we have found 45 independent amplitudes.
Similarly, we have another 45 amplitudes for the color 
structure $\frac{1}{3}\delta_{il}\delta_{jk}$. Together,
we have 90 independent amplitudes for the $u\bar d q\bar q$
component. 

The above formalism can also be used to
construct the amplitudes for the $u\overline d s\overline s$
and $u\overline d c\overline c$ components in $\pi^+$.
The total number of independent amplitudes are 
90 in both cases. These amplitudes can be used to describe 
the intrinsic strange and/or charm contributions to the 
hadronic processes involving $\pi$, e.g., $J/\psi\rightarrow 
\rho\pi$ decays \cite{Brodsky:1997fj}.

\section{Wave-function Amplitudes for the $\rho^+$-meson}

The method in the last section can be straightforwardly used
to construct the light-cone wave-function amplitudes for the 
$\rho$ mesons. Strictly speaking, the $\rho$ meson is
not an eigenstate of the QCD hamiltonian, it appears as
resonances in, for example, $\pi\pi$ scattering. However,
we regard in the following discussion the $\rho$ meson
as if a bound state of quarks and gluons.
The relevant studies of the distribution amplitudes 
for $\rho$ mesons can be found in 
\cite{Chernyak:1984bm,Ball:1998sk}.

Because $\rho$ is a vector meson, it has three helicity states,
i.e., $\Lambda=0, \pm 1$, corresponding to
longitudinal ($\Lambda=0$) and transverse ($\Lambda=\pm 1$)
polarizations. The wave functions for the $\Lambda=0$ state 
can be obtained, in principle, from those of the $\Lambda=\pm 1$ states
by using angular momentum raising and lowering operators. In
practice, however, these operators involve complicated quark-gluon
interactions in light-cone quantization, and the constraint
becomes a very complicated equation involving all higher
Fock states. Since we are interested in the components
of few partons, we may regard the different helicity states
as quasi-independent. Nonetheless, the $\Lambda=-1$ state can be obtained 
from $\Lambda=+1$ state using parity transformation.

The wave-function amplitudes for the helicity $\Lambda=0$ 
state can be easily obtained from those in the last section, taking
into account the difference on $\hat Y$ transformation property
between $\pi$ and $\rho$, i.e.,
\begin{equation}
\hat Y \vert \rho^+,\Lambda=0 \rangle =\vert \rho^+, \Lambda=0\rangle \ ,
\label{rhoy}
\end{equation}
compared to Eq. (\ref{piy}).
Hence the Fock expansion listed in the last section can be 
transformed to that of $|\rho^+,0\rangle$, 
except some signs must be changed. In particular, the total number of 
the independent amplitudes will be the same.

For example, the two quark component for $|\rho^+,0\rangle$ has 2 independent
amplitudes 
\begin{eqnarray}
  |\rho^+,0\rangle^{l_z=0}_{u\overline d} &=& \int d[1]d[2] \psi_{u\overline d}^{(1)}(1,2)  
\frac{1}{\sqrt{3}}\left[u^\dagger_{\uparrow i}(1)
  \overline d^\dagger_{\downarrow i}(2) +
u^\dagger_{\downarrow i}(1)
\overline d^\dagger_{\uparrow i}(2)\right]
 |0\rangle \, \\ 
  |\rho^+,0\rangle _{u\overline d}^{|l_z|=1}&=& \int d[1]d[2] \psi_{u\overline d}^{(2)}(1,2)  
\frac{1}{\sqrt{3}}\left[ k_{1\perp}^-u^\dagger_{\uparrow i}(1)
  \overline d^\dagger_{\uparrow i}(2)
- k_{1\perp}^+u^\dagger_{\downarrow i}(1)
  \overline d^\dagger_{\downarrow i}(2)\right] |0\rangle \ .
\end{eqnarray}
Here we have used the same notation for the amplitudes, assuming no confusion
will arise. For the $u\bar d g$ component, we have
\begin{eqnarray}
  |\rho^+,0\rangle^{l_z=0}_{u\overline d g} &=& \int d[1]d[2]d[3] 
\frac{T^a_{ij}}{2}
\left\{\psi_{u\overline d g}^{(1)}(1,2,3)
\left[(u\overline d)^\dagger_{A,1}g^{a\dagger}_\downarrow (3)+(u\overline d)^\dagger_{A,-1}g^{a\dagger}_\uparrow (3)\right]
|0\rangle\right.\nonumber\\
&&\left.+i\epsilon^{\alpha\beta}k_{1\alpha}k_{2\beta}\psi_{u\overline d g}^{(2)}(1,2,3)
\left[(u\overline d)^\dagger_{A,1}g^{a\dagger}_\downarrow (3)-(u\overline d)^\dagger_{A,-1}g^{a\dagger}_\uparrow (3)\right]
\right\}|0\rangle \ ,\\
  |\rho^+,0\rangle^{|l_z|=1}_{u\overline d g} &=& \int d[1]d[2]d[3] 
\frac{T^a_{ij}}{2}
\left\{
 \pn3 ^{(3)}(1,2,3)
	 \left[k_{1\perp}^+ (u\overline d)^\dagger_{A,0}g^{a\dagger}_\downarrow (3) +k_{1\perp}^-
   (u\overline d)^\dagger_{A,0}g^{a\dagger}_\uparrow (3)\right] \right. \nonumber\\
   && +\pn3 ^{(4)}(1,2,3) 
	\left[k_{2\perp}^+ (u\overline d)^\dagger_{A,0}g^{a\dagger}_\downarrow (3) +k_{2\perp}^-
   (u\overline d)^\dagger_{A,0}g^{a\dagger}_\uparrow (3)\right] \nonumber\\
   && +\pn3 ^{(5)}(1,2,3) 
	\left[k_{1\perp}^+ (u\overline d)^\dagger_{S,0}g^{a\dagger}_\downarrow (3) -k_{1\perp}^-
   (u\overline d)^\dagger_{S,0}g^{a\dagger}_\uparrow (3)\right] \nonumber\\
   && \left.+\pn3 ^{(6)}(1,2,3) 
	\left[k_{2\perp}^+ (u\overline d)^\dagger_{S,0}g^{a\dagger}_\downarrow (3) -k_{2\perp}^-
   (u\overline d)^\dagger_{S,0}g^{a\dagger}_\uparrow (3)\right] 
\right\}
 |0\rangle \ ,
\end{eqnarray}
\begin{eqnarray}
  |\rho^+,0\rangle^{|l_z|=2}_{u\overline d g} &=& \int d[1]d[2]d[3] 
\frac{T^a_{ij}}{2}
\left\{
 \pn3 ^{(7)} (1,2,3)
   \left[k_{1\perp}^+k_{1\perp}^+ (u\overline d)^\dagger_{A,-1}g^{a\dagger}_\downarrow (3)+ k_{1\perp}^-k_{1\perp}^- 
(u\overline d)^\dagger_{A,1}g^{a\dagger}_\uparrow (3)\right]\right.
   \nonumber\\
  && + \pn3 ^{(8)} (1,2,3)
   \left[k_{1\perp}^+k_{2\perp}^+ (u\overline d)^\dagger_{A,-1}g^{a\dagger}_\downarrow (3)+ k_{1\perp}^-k_{2\perp}^- 
(u\overline d)^\dagger_{A,1}g^{a\dagger}_\uparrow (3)\right]
  \nonumber\\
  && \left.+\pn3 ^{(9)}(1,2,3)
   \left[k_{2\perp}^+k_{2\perp}^+ (u\overline d)^\dagger_{A,-1}g^{a\dagger}_\downarrow (3)+ k_{2\perp}^-k_{2\perp}^- 
(u\overline d)^\dagger_{A,1}g^{a\dagger}_\uparrow (3)\right]\right\}
|0\rangle \ .
\end{eqnarray}
We will not repeat the cases for four partons. 

The helicity $\Lambda=1$ $\rho$-meson state, $|\rho^+,1\rangle$, 
can be constructed similarly. 
For example, the $u \bar d$ component defined 4 independent amplitudes,
corresponding to $l_z=0,\ 1,\ 2$,
\begin{eqnarray}
  |\rho^+,1\rangle^{l_z=0}_{u\overline d} &=& \int d[1]d[2] \psi_{u\overline d}^{(1)}(1,2)  
\frac{1}{\sqrt{3}}\left[u^\dagger_{\uparrow i}(1)
  \overline d^\dagger_{\uparrow i}(2) \right]
 |0\rangle \, \\ 
  |\rho^+,1\rangle^{l_z=1}_{u\overline d} &=& \int d[1]d[2] 
\left\{k_{1\perp}^+\psi_{u\overline d}^{(2)}(1,2)  
\frac{1}{\sqrt{3}}\left[u^\dagger_{\uparrow i}(1)
  \overline d^\dagger_{\downarrow i}(2) +
u^\dagger_{\downarrow i}(1)
\overline d^\dagger_{\uparrow i}(2)\right]\right.\nonumber\\
&&\left.+k_{1\perp}^+\psi_{u\overline d}^{(3)}(1,2)  
\frac{1}{\sqrt{3}}\left[u^\dagger_{\uparrow i}(1)
  \overline d^\dagger_{\downarrow i}(2) -
u^\dagger_{\downarrow i}(1)
\overline d^\dagger_{\uparrow i}(2)\right]
\right\}
 |0\rangle \, \\ 
  |\rho^+,1\rangle _{u\overline d}^{l_z=2}&=& \int d[1]d[2] 
(k_{1\perp}^+)^2\psi_{u\overline d}^{(4)}(1,2)  
\frac{1}{\sqrt{3}}
 \left[u^\dagger_{\downarrow i}(1)
  \overline d^\dagger_{\downarrow i}(2)\right] |0\rangle \ .
\end{eqnarray}
Here again we use the same notation for the wave-function amplitudes 
although they can be very different from those in the $\Lambda=0$ state. 
In fact, even the number of independent amplitudes for a given number of
partons is different. 

For the $u\bar d g$ component, the parton orbital angular momentum 
can be $l_z=-1, 0, 1, 2,$ and 3.
For $l_z=0$, we find 4 independent amplitudes
\begin{eqnarray}
  |\rho^+,1\rangle^{l_z=0}_{u\overline d g} &=& \int d[1]d[2]d[3] 
\frac{T^a_{ij}}{2}
\left\{\left(\psi_{u\overline d g}^{(1)}(1,2,3)
+i\epsilon^{\alpha\beta}k_{1\alpha}k_{2\beta}\psi_{u\overline d g}^{(2)}(1,2,3)\right)
\left[(u\overline d)^\dagger_{A,0}g^{a\dagger}_\uparrow (3)\right]\right.\nonumber\\
&&\left.+\left(\psi_{u\overline d g}^{(3)}(1,2,3)
+i\epsilon^{\alpha\beta}k_{1\alpha}k_{2\beta}\psi_{u\overline d g}^{(4)}(1,2,3)\right)
\left[(u\overline d)^\dagger_{S,0}g^{a\dagger}_\uparrow (3)\right]
\right\}|0\rangle \ .
\end{eqnarray}
For $l_z=-1$, we have 2 independent amplitudes
\begin{eqnarray}
  |\rho^+,1\rangle^{l_z=-1}_{u\overline d g} &=& \int d[1]d[2]d[3] 
\frac{T^a_{ij}}{2}
\left\{
 \pn3 ^{(5)}(1,2,3)
	 \left[k_{1\perp}^- (u\overline d)^\dagger_{A,1}g^{a\dagger}_\uparrow (3) \right] 
\right. \nonumber\\
   && \left. +\pn3 ^{(6)}(1,2,3) 
	\left[k_{2\perp}^- (u\overline d)^\dagger_{A,1}g^{a\dagger}_\uparrow (3) \right] 
\right\}
 |0\rangle \ .
\end{eqnarray}
For $l_z=1$, we have 4 independent amplitudes
\begin{eqnarray}
  |\rho^+,1\rangle^{l_z=1}_{u\overline d g} &=& \int d[1]d[2]d[3] 
\frac{T^a_{ij}}{2}
\left\{
k_{1\perp}^+ \pn3 ^{(7)}(1,2,3)
	 \left[(u\overline d)^\dagger_{A,1}g^{a\dagger}_\downarrow (3) +
   (u\overline d)^\dagger_{A,-1}g^{a\dagger}_\uparrow (3)\right] \right. \nonumber\\
   && +k_{2\perp}^+\pn3 ^{(8)}(1,2,3) 
	 \left[(u\overline d)^\dagger_{A,1}g^{a\dagger}_\downarrow (3) +
   (u\overline d)^\dagger_{A,-1}g^{a\dagger}_\uparrow (3)\right] \nonumber\\
   && +k_{1\perp}^+\pn3 ^{(9)}(1,2,3) 
		 \left[(u\overline d)^\dagger_{A,1}g^{a\dagger}_\downarrow (3) -
   (u\overline d)^\dagger_{A,-1}g^{a\dagger}_\uparrow (3)\right] \nonumber\\
   && \left.+k_{2\perp}^+\pn3 ^{(10)}(1,2,3) 
	 \left[(u\overline d)^\dagger_{A,1}g^{a\dagger}_\downarrow (3) -
   (u\overline d)^\dagger_{A,-1}g^{a\dagger}_\uparrow (3)\right] 
\right\}
 |0\rangle \ .
\end{eqnarray}
For $l_z=2$, we have 6 independent amplitudes
\begin{eqnarray}
  |\rho^+,1\rangle^{l_z=2}_{u\overline d g} &=& \int d[1]d[2]d[3] 
\frac{T^a_{ij}}{2}
\left\{
 k_{1\perp}^+k_{1\perp}^+ \pn3 ^{(11)} (1,2,3)
   \left[(u\overline d)^\dagger_{A,0}g^{a\dagger}_\downarrow (3)\right]\right.
   \nonumber\\
  && + k_{1\perp}^+k_{2\perp}^+\pn3 ^{(12)} (1,2,3)
   \left[ (u\overline d)^\dagger_{A,0}g^{a\dagger}_\downarrow (3)\right]
  \nonumber\\
  && +k_{2\perp}^+k_{2\perp}^+\pn3 ^{(13)}(1,2,3)
   \left[ (u\overline d)^\dagger_{A,0}g^{a\dagger}_\downarrow (3)\right]
  \nonumber\\
 &&+k_{1\perp}^+k_{1\perp}^+ \pn3 ^{(14)} (1,2,3)
   \left[ (u\overline d)^\dagger_{S,0}g^{a\dagger}_\downarrow (3)\right]
   \nonumber\\
  && + k_{1\perp}^+k_{2\perp}^+\pn3 ^{(15)} (1,2,3)
   \left[ (u\overline d)^\dagger_{S,0}g^{a\dagger}_\downarrow (3)\right]
  \nonumber\\
  &&\left. +k_{2\perp}^+k_{2\perp}^+\pn3 ^{(16)}(1,2,3)
   \left[ (u\overline d)^\dagger_{S,0}g^{a\dagger}_\downarrow (3)\right]
\right\}
|0\rangle \ .
\end{eqnarray}
For $l_z=3$, we have 4 independent amplitudes,
\begin{eqnarray}
  |\rho^+,1\rangle^{l_z=3}_{u\overline d g} &=& \int d[1]d[2]d[3] 
\frac{T^a_{ij}}{2}
\left\{
 (k_{1\perp}^+)^3\pn3 ^{(17)} (1,2,3)
   \left[(u\overline d)^\dagger_{A,-1}g^{a\dagger}_\downarrow (3)\right]\right.
   \nonumber\\
  && +(k_{2\perp}^+)^3\pn3 ^{(18)}(1,2,3)
   \left[ (u\overline d)^\dagger_{A,-1}g^{a\dagger}_\downarrow (3)\right]
  \nonumber\\
  && + (k_{1\perp}^+)^2k_{2\perp}^+\pn3 ^{(19)} (1,2,3)
   \left[ (u\overline d)^\dagger_{A,-1}g^{a\dagger}_\downarrow (3)\right]
  \nonumber\\
  &&\left. +k_{1\perp}^+(k_{2\perp}^+)^2\pn3 ^{(20)}(1,2,3)
   \left[ (u\overline d)^\dagger_{A,-1}g^{a\dagger}_\downarrow (3)\right]
\right\}
|0\rangle \ .
\end{eqnarray}
In total, we have 20 amplitudes, compared with the $\Lambda=0$ case where
we have only 9. For simplicity, we will not consider
those amplitudes with four partons. 

\section{Wave-function Amplitudes for the Nucleon}

In this section, we enumerate the number of independent amplitudes
for the nucleon, and more specifically for the proton. For the
neutron, one just interchange the up and down quarks assuming 
isospin symmetry. Our expansion
is also valid for the whole baryon octet, except the flavor structure
need to be modified accordingly.

We consider only the state with positive helicity. The negative
helicity state can be obtained simply from the modified
parity transformation $\hat Y$. Three quark amplitudes have been
studied extensively in Ref.~\cite{Ji:2002xn}. The new result 
here includes three quark plus one gluon amplitudes. One
can add an additional pair of sea quarks into the 
valence component, but the result is very complicated and
we will not show it here.

\subsection{The $uud$ Component}

The quark distribution amplitudes describing the three-quark 
component of the proton have been studied extensively in the 
literature \cite{Lepage:1980fj,Chernyak:1984bm,King:1987wi,Chernyak:1989nv,
Braun:1999te,Braun:2000kw}. 
The wave-function amplitudes keeping full partons
transverse-momentum dependence have been 
studied in Ref.~\cite{Ji:2002xn}. From the approach advocated here,
we immediate have for $l_z=0$ and $l_z=1$, 
\begin{eqnarray}
  |P\uparrow\rangle_{uud}^{l_z=0} &=& \int d[1]d[2]d[3]\left( \psi_{uud}^{(1)}(1,2,3)
         + i\epsilon^{\alpha\beta}k_{1\alpha}k_{2\beta}  \psi_{uud}^{(2)}(1,2,3)\right) \nonumber \\
         &&  \times  \frac{\epsilon^{ijk}}{\sqrt{6}} u^{\dagger}_{i\uparrow}(1)
             \left(u^{\dagger}_{j\downarrow}(2)d^{\dagger}_{k\uparrow}(3)
            -d^{\dagger}_{j\downarrow}(2)u^{\dagger}_{k\uparrow}(3)\right)
         |0\rangle \ , \\
  |P\uparrow\rangle_{uud}^{l_z=1} &=& \int d[1]d[2]d[3]\left(k_{1\perp}^+
           \psi_{uud}^{(3)}(1,2,3)
         + k_{2\perp}^+ \psi_{uud}^{(4)}(1,2,3)\right) \nonumber \\
         &&  \times  \frac{\epsilon^{ijk}}{\sqrt{6}} \left( u^{\dagger}_{i\uparrow}(1)
            u^{\dagger}_{j\downarrow}(2)d^{\dagger}_{k\downarrow}(3)
            -d^{\dagger}_{i\uparrow}(1)u^{\dagger}_{j\downarrow}(2)
             u^{\dagger}_{k\downarrow}(3)\right)
         |0\rangle \ .
\end{eqnarray}
For $l_z=-1$, we have 
\begin{eqnarray}
  |P\uparrow\rangle_{uud}^{l_z=-1} &=& \int d[1]d[2]d[3]~k_{2\perp}^-
          \psi_{uud}^{(5)}(1,2,3) \nonumber \\
         &&  \times  \frac{\epsilon^{ijk}}{\sqrt{6}} u^{\dagger}_{i\uparrow}(1)
             \left(
     u^{\dagger}_{j\uparrow}(2)d^{\dagger}_{k\uparrow}(3)
    -d^{\dagger}_{j\uparrow}(2)u^{\dagger}_{k\uparrow}(3)
            \right)
         |0\rangle \ ,
\end{eqnarray}
where we have used quark 2 and 3 anti-symmetry.
Likewise, we have,
\begin{eqnarray}
  |P\uparrow\rangle_{uud}^{l_z=2} &=& \int d[1]d[2]d[3]~k_{1\perp}^+k_{3\perp}^+
         \psi_{uud}^{(6)}(1,2,3)  \nonumber \\
         &&  \times  \frac{\epsilon^{ijk}}{\sqrt{6}} u^{\dagger}_{i\downarrow}(1)
             \left(d^{\dagger}_{j\downarrow}(2)u^{\dagger}_{k\downarrow}(3)-u^{\dagger}_{j\downarrow}(2)d^
{\dagger}_{k\downarrow}(3)
            \right)
         |0\rangle \ .
\end{eqnarray}
where, in principle, there is an additional term $k_{2\perp}^+k_{3\perp}^+
\psi_{uud}^{(6')}(1,2,3)$. However, after using 2 and 3 anti-symmetry 
and 1 and 2 symmetry, it can shown that this term can be reduced to 
the term shown above. 
   
\subsection{The $uud+g$ Component}

Let us first consider the isospin symmetry. With three quarks $uud$,
one can construct two possible $I=1/2$ isospin 
combinations:
\begin{eqnarray}
&&u(ud-du)\ ,\nonumber\\
&&2duu-udu-uud\ .
\end{eqnarray}
However, the second flavor structure can be reduced to
the first one after some shuffling of the particle labels.
Therefore, we shall only consider the first structure
for the flavor wave function, taking into account all possible 
color and spin assignments for the three quarks and one gluon.

Since the gluon belongs to color-octet,
the three quarks must couple to a color-octet.
For the three quarks, $u_i,~u_j,~d_k$, there are two 
possible ways to couple to color-octet 
$$3\times 3\times 3=(6+\bar 3)\times 3=1+8+8+10 \ . $$
We can have the first two quarks couple to $\bar 3$, and then
couple them to the third quark to form a color-octet. In this
case, we have an overall color factor, $\epsilon^{ijl}T^a_{lk}$.
Similarly, we can also have other two color factors,
$\epsilon^{jkl}T^a_{li}$ and $\epsilon^{kil}T^a_{lj}$.
However, the above three are not independent, because 
$\epsilon^{ijl}T^a_{lk}+\epsilon^{jkl}T^a_{li}+\epsilon^{kil}T^a_{lj}= 0 $.
If we use the isospin structure $u_i(u_jd_k-d_ju_k)$, the best
way to select the two independent color structures 
is to have the indices of $jk$ to be antisymmetric or 
symmetric:
\begin{equation}
\epsilon^{jkl}T^a_{li} \ , ~~~~~ \epsilon^{ijl}T^a_{lk}+
\epsilon^{ikl}T^a_{lj} \ . 
\end{equation}

For the Fock component of $uudg$, the total quark helicity can be
$\lambda_{uud}=-3/2,~-1/2,~1/2,~3/2$, and the gluon helicity 
$\lambda_{g}=\pm 1$. The parton orbital angular momentum projection can
have the following values, $l_z=0, 1, 2, 3, -1, -2$.

For $l_z=0$, the parton helicity can either be $\lambda_{uud}=3/2$ and $\lambda_g=-1$, or 
$\lambda_{uud}=-1/2$ and $\lambda_g=1$. For the first case, because
the total quark helicity $\lambda_{uud}=3/2$, the three quarks are all
in helicity-1/2 states, and we only have one spin structure, i.e.,
	$u_{i\uparrow}(u_{j\uparrow}d_{k\uparrow}-d_{j\uparrow}u_{k\uparrow})$.
Therefore, we can write down 3 independent amplitudes,
\begin{eqnarray}
  |P\uparrow\rangle_{uudg}^{l_z=0} &=& \int d[1]d[2]d[3]d[4]\nonumber\\
&&	\left( \psi_{uudg}^{(1)}(1,2,3,4)
+ i\epsilon^{\alpha\beta}k_{1\alpha}k_{2\beta}  \psi_{uudg}^{(2)}(1,2,3,4)
         + i\epsilon^{\alpha\beta}k_{2\alpha}k_{3\beta}  \psi_{uudg}^{(3)}(1,2,3,4)\right) \nonumber \\
         &&  \times \frac{\epsilon^{jkl}T^a_{li}}{2}\left[  u^{\dagger}_{i\uparrow}(1)
             \left(u^{\dagger}_{j\uparrow}(2)d^{\dagger}_{k\uparrow}(3)
            -d^{\dagger}_{j\uparrow}(2)u^{\dagger}_{k\uparrow}(3)\right)
	g^{a\dagger}_{\downarrow}(4)
	\right]
         |0\rangle \ .
\end{eqnarray}
Here, we have used the $2\leftrightarrow 3$ symmetry to reduce the number of
the independent amplitudes. For example, 
$i\epsilon^{\alpha\beta}k_{1\alpha}k_{3\beta}$ term can be obtained 
from $i\epsilon^{\alpha\beta}k_{1\alpha}k_{2\beta}$ by 2 and 3 exchange,
and hence the former is not independent. 
We have the following (anti)symmetric properties for some amplitudes,
$\psi_{uudg}^{(1)}(1,2,3,4)=-\psi_{uudg}^{(1)}(1,3,2,4)$
and $\psi_{uudg}^{(3)}(1,2,3,4)=\psi_{uudg}^{(3)}(1,3,2,4)$.

In the second case, the total quark helicity $\lambda_{uud}=-1/2$. 
There are three possible spin structures for the three quarks,
\begin{eqnarray}
&&u_{i\downarrow}(1)\left(u_{j\downarrow}(2)d_{k\uparrow}(3)-d_{j\downarrow}(2)u_{k\uparrow}(3)\right) \ , \nonumber\\
&&u_{i\downarrow}(1)\left(u_{j\uparrow}(2)d_{k\downarrow}(3)-d_{j\uparrow}(2)u_{k\downarrow}(3)\right)\ , \nonumber\\
&&u_{i\uparrow}(1)\left(u_{j\downarrow}(2)d_{k\downarrow}(3)-d_{j\downarrow}(2)u_{k\downarrow}(3)\right)\ .
\end{eqnarray}
However, the associated color structures indicate that there is a (anti)symmetric
relation between the two indices $j$ and $k$. Thus,
the first two spin structures are equivalent to each other
under 2 and 3 exchange. In the following, we will only keep one of the
first two spin structures. The above observation also applies to 
the total quark helicity $\lambda_{uud}=1/2$ case. 
Taking into account this, we find 7 independent amplitudes
\begin{eqnarray}
  |P\uparrow\rangle_{uudg}^{l_z=0} &=& \int d[1]d[2]d[3]d[4]
	\left\{\left( \psi_{uudg}^{(4)}(1,2,3,4)
         + i\epsilon^{\alpha\beta}k_{1\alpha}k_{2\beta}  \psi_{uudg}^{(5)}(1,2,3,4)
\right. \right. \nonumber \\
     && \left. + i\epsilon^{\alpha\beta}k_{1\alpha}k_{3\beta}  \psi_{uudg}^{(6)}(1,2,3,4)
	+ i\epsilon^{\alpha\beta}k_{2\alpha}k_{3\beta}  \psi_{uudg}^{(7)}(1,2,3,4)
\right) \nonumber \\
  &&  \times \frac{\epsilon^{jkl}T^a_{li}}{2}\left[  u^{\dagger}_{i\downarrow}(1)
             \left(u^{\dagger}_{j\uparrow}(2)d^{\dagger}_{k\downarrow}(3)
            -d^{\dagger}_{j\uparrow}(2)u^{\dagger}_{k\downarrow}(3)\right)
	g^{a\dagger}_{\uparrow}(4)
	\right]\nonumber\\
&&+\left( \psi_{uudg}^{(8)}(1,2,3,4)
         + i\epsilon^{\alpha\beta}k_{1\alpha}k_{2\beta}  \psi_{uudg}^{(9)}(1,2,3,4)
	+ i\epsilon^{\alpha\beta}k_{2\alpha}k_{3\beta}  \psi_{uudg}^{(10)}(1,2,3,4)
\right) \nonumber \\
  &&\left.  \times \frac{\epsilon^{jkl}T^a_{li}}{2}\left[  u^{\dagger}_{i\uparrow}(1)
             \left(u^{\dagger}_{j\downarrow}(2)d^{\dagger}_{k\downarrow}(3)
            -d^{\dagger}_{j\downarrow}(2)u^{\dagger}_{k\downarrow}(3)\right)
	g^{a\dagger}_{\uparrow}(4)
	\right]\right\}
         |0\rangle \ .
\end{eqnarray}
Again, the 2 and 3 symmetry in the above equation
has been used to reduce the number of independent amplitudes.
Moreover, it implies the relations $\psi_{uudg}^{(8)}(1,2,3,4)=-\psi_{uudg}^{(8)}(1,3,2,4)$
and $\psi_{uudg}^{(10)}(1,2,3,4)=\psi_{uudg}^{(10)}(1,3,2,4)$.

For $l_z=1$, the parton helicity can either be $\lambda_{uud}=1/2$ and $\lambda_g=-1$, or 
$\lambda_{uud}=-3/2$ and $\lambda_g=1$. In the first case, we define 10
independent amplitudes
\begin{eqnarray}
  |P\uparrow\rangle_{uudg}^{l_z=1} &=& \int d[1]d[2]d[3]d[4]
	\left\{\left( k_{1\perp}^+(\psi_{uudg}^{(11)}(1,2,3,4)
         + i\epsilon^{\alpha\beta}k_{2\alpha}k_{3\beta}  \psi_{uudg}^{(12)}(1,2,3,4))
\right. \right.\nonumber \\
&&+ k_{2\perp}^+(\psi_{uudg}^{(13)}(1,2,3,4)
         + i\epsilon^{\alpha\beta}k_{1\alpha}k_{3\beta}  \psi_{uudg}^{(14)}(1,2,3,4))
\nonumber \\
     && \left. 
+ k_{3\perp}^+(\psi_{uudg}^{(15)}(1,2,3,4)
         + i\epsilon^{\alpha\beta}k_{1\alpha}k_{2\beta}  \psi_{uudg}^{(16)}(1,2,3,4))
\right) \nonumber \\
  &&  \times \frac{\epsilon^{jkl}T^a_{li}}{2}\left[  u^{\dagger}_{i\uparrow}(1)
             \left(u^{\dagger}_{j\uparrow}(2)d^{\dagger}_{k\downarrow}(3)
            -d^{\dagger}_{j\uparrow}(2)u^{\dagger}_{k\downarrow}(3)\right)
	g^{a\dagger}_{\downarrow}(4)
	\right]\nonumber\\
&&+\left( k_{1\perp}^+(\psi_{uudg}^{(17)}(1,2,3,4)
         + i\epsilon^{\alpha\beta}k_{2\alpha}k_{3\beta}  \psi_{uudg}^{(18)}(1,2,3,4))
\right. \nonumber \\
&&\left.+ k_{2\perp}^+(\psi_{uudg}^{(19)}(1,2,3,4)
         + i\epsilon^{\alpha\beta}k_{1\alpha}k_{3\beta}  \psi_{uudg}^{(20)}(1,2,3,4))
\right) \nonumber \\
  && \left. \times \frac{\epsilon^{jkl}T^a_{li}}{2}\left[  u^{\dagger}_{i\downarrow}(1)
             \left(u^{\dagger}_{j\uparrow}(2)d^{\dagger}_{k\uparrow}(3)
            -d^{\dagger}_{j\uparrow}(2)u^{\dagger}_{k\uparrow}(3)\right)
	g^{a\dagger}_{\downarrow}(4)
	\right]\right\}
         |0\rangle \ .
\end{eqnarray}
The symmetry between 2 and 3 leads to $\psi_{uudg}^{(17)}(1,2,3,4)
=-\psi_{uudg}^{(17)}(1,3,2,4)$
and $\psi_{uudg}^{(18)}(1,2,3,4)=\psi_{uudg}^{(18)}(1,3,2,4)$.
In the second case, we define 4 independent amplitudes
\begin{eqnarray}
  |P\uparrow\rangle_{uudg}^{l_z=1} &=& \int d[1]d[2]d[3]d[4]
	\left(
	 k_{1\perp}^+(\psi_{uudg}^{(21)}(1,2,3,4)
         + i\epsilon^{\alpha\beta}k_{2\alpha}k_{3\beta}  \psi_{uudg}^{(22)}(1,2,3,4))
 \right.\nonumber \\
&&+\left.
k_{2\perp}^+(\psi_{uudg}^{(23)}(1,2,3,4)
         + i\epsilon^{\alpha\beta}k_{1\alpha}k_{3\beta}  \psi_{uudg}^{(24)}(1,2,3,4))
	\right)\nonumber\\
  &&  \times \frac{\epsilon^{jkl}T^a_{li}}{2}\left[  u^{\dagger}_{i\downarrow}(1)
             \left(u^{\dagger}_{j\downarrow}(2)d^{\dagger}_{k\downarrow}(3)
            -d^{\dagger}_{j\downarrow}(2)u^{\dagger}_{k\downarrow}(3)\right)
	g^{a\dagger}_{\uparrow}(4)
	\right]
         |0\rangle \ ,
\end{eqnarray}
where $\psi_{uudg}^{(21)}(1,2,3,4)=-\psi_{uudg}^{(21)}(1,3,2,4)$
and $\psi_{uudg}^{(22)}(1,2,3,4)=\psi_{uudg}^{(22)}(1,3,2,4)$.

For $l_z=2$, the parton helicity must be 
$\lambda_{uud}=-1/2$ and $\lambda_g=-1$. We define 15 independent amplitudes
\begin{eqnarray}
  |P\uparrow\rangle_{uudg}^{l_z=2} &=& \int d[1]d[2]d[3]d[4]
	\left\{\left( (k_{1\perp}^+)^2(\psi_{uudg}^{(25)}(1,2,3,4)
         + i\epsilon^{\alpha\beta}k_{2\alpha}k_{3\beta}  \psi_{uudg}^{(26)}(1,2,3,4))
\right.\right. \nonumber \\
&&+ (k_{2\perp}^+)^2(\psi_{uudg}^{(27)}(1,2,3,4)
         + i\epsilon^{\alpha\beta}k_{1\alpha}k_{3\beta}  \psi_{uudg}^{(28)}(1,2,3,4))
\nonumber \\
          &&+ (k_{3\perp}^+)^2(\psi_{uudg}^{(29)}(1,2,3,4)
         + i\epsilon^{\alpha\beta}k_{1\alpha}k_{2\beta}  \psi_{uudg}^{(30)}(1,2,3,4))
 \nonumber \\
     && \left. 
+  k_{1\perp}^+k_{2\perp}^+\psi_{uudg}^{(31)}(1,2,3,4)+k_{1\perp}^+k_{3\perp}^+\psi_{uudg}^{(32)}(1,2,3,4)
+k_{2\perp}^+k_{3\perp}^+\psi_{uudg}^{(33)}(1,2,3,4)
\right) \nonumber \\
  &&  \times \frac{\epsilon^{jkl}T^a_{li}}{2}\left[  u^{\dagger}_{i\downarrow}(1)
             \left(u^{\dagger}_{j\uparrow}(2)d^{\dagger}_{k\downarrow}(3)
            -d^{\dagger}_{j\uparrow}(2)u^{\dagger}_{k\downarrow}(3)\right)
	g^{a\dagger}_{\downarrow}(4)
	\right]\nonumber\\
&&+\left( (k_{1\perp}^+)^2(\psi_{uudg}^{(34)}(1,2,3,4)
         + i\epsilon^{\alpha\beta}k_{2\alpha}k_{3\beta}  \psi_{uudg}^{(35)}(1,2,3,4))
\right. \nonumber \\
&&+ (k_{2\perp}^+)^2(\psi_{uudg}^{(36)}(1,2,3,4)
         + i\epsilon^{\alpha\beta}k_{1\alpha}k_{3\beta}  \psi_{uudg}^{(37)}(1,2,3,4))
\nonumber
\end{eqnarray}
\begin{eqnarray}
     && \left. 
+  k_{1\perp}^+k_{2\perp}^+\psi_{uudg}^{(38)}(1,2,3,4)
+k_{2\perp}^+k_{3\perp}^+\psi_{uudg}^{(39)}(1,2,3,4)
\right) \nonumber \\
  && \left. \times \frac{\epsilon^{jkl}T^a_{li}}{2}\left[  u^{\dagger}_{i\uparrow}(1)
             \left(u^{\dagger}_{j\downarrow}(2)d^{\dagger}_{k\downarrow}(3)
            -d^{\dagger}_{j\downarrow}(2)u^{\dagger}_{k\downarrow}(3)\right)
	g^{a\dagger}_{\downarrow}(4)
	\right]\right\}
         |0\rangle \ ,
\end{eqnarray}
where $\psi_{uudg}^{(34,39)}(1,2,3,4)=-\psi_{uudg}^{(34,39)}(1,3,2,4)$
and $\psi_{uudg}^{(35)}(1,2,3,4)=\psi_{uudg}^{(35)}(1,3,2,4)$.

For $l_z=3$, the parton helicity must be $\lambda_{uud}=-3/2$ and $\lambda_g=-1$. 
We define 8 independent amplitudes:
\begin{eqnarray}
  |P\uparrow\rangle_{uudg}^{l_z=3} &=& \int d[1]d[2]d[3]d[4]
	\left( (k_{1\perp}^+)^3(\psi_{uudg}^{(40)}(1,2,3,4)
         + i\epsilon^{\alpha\beta}k_{2\alpha}k_{3\beta}  \psi_{uudg}^{(41)}(1,2,3,4))
\right. \nonumber \\
     &&+(k_{2\perp}^+)^3(\psi_{uudg}^{(42)}(1,2,3,4)
         + i\epsilon^{\alpha\beta}k_{1\alpha}k_{3\beta}  \psi_{uudg}^{(43)}(1,2,3,4))
 \nonumber \\
&&+  (k_{1\perp}^+)^2k_{2\perp}^+\psi_{uudg}^{(44)}(1,2,3,4)+(k_{2\perp}^+)^2k_{1\perp}^+\psi_{uudg}^{(45)}(1,2,3,4)
\nonumber \\
&& \left. 
+  (k_{2\perp}^+)^2k_{3\perp}^+\psi_{uudg}^{(46)}(1,2,3,4)+k_{1\perp}^+k_{2\perp}^+k_{3\perp}^+\psi_{uudg}^{(47)}(1,2,3,4)
\right) \nonumber \\
  &&  \times \frac{\epsilon^{jkl}T^a_{li}}{2}\left[  u^{\dagger}_{i\downarrow}(1)
             \left(u^{\dagger}_{j\downarrow}(2)d^{\dagger}_{k\downarrow}(3)
            -d^{\dagger}_{j\downarrow}(2)u^{\dagger}_{k\downarrow}(3)\right)
	g^{a\dagger}_{\downarrow}(4)
	\right]
         |0\rangle \ ,
\end{eqnarray}
where $\psi_{uudg}^{(40,47)}(1,2,3,4)=-\psi_{uudg}^{(40,47)}(1,3,2,4)$
and $\psi_{uudg}^{(41)}(1,2,3,4)=\psi_{uudg}^{(41)}(1,3,2,4)$.

For $l_z=-1$, the parton helicity must be $\lambda_{uud}=1/2$ and $\lambda_g=1$.
In this case, we define 10 independent amplitudes  
\begin{eqnarray}
  |P\uparrow\rangle_{uudg}^{l_z=-1} &=& \int d[1]d[2]d[3]d[4]
	\left\{\left( k_{1\perp}^-(\psi_{uudg}^{(48)}(1,2,3,4)
         + i\epsilon^{\alpha\beta}k_{2\alpha}k_{3\beta}  \psi_{uudg}^{(49)}(1,2,3,4))
\right.\right. \nonumber \\
&&+ k_{2\perp}^-(\psi_{uudg}^{(50)}(1,2,3,4)
         + i\epsilon^{\alpha\beta}k_{1\alpha}k_{3\beta}  \psi_{uudg}^{(51)}(1,2,3,4))
\nonumber \\
     && \left. 
+ k_{3\perp}^-(\psi_{uudg}^{(52)}(1,2,3,4)
         + i\epsilon^{\alpha\beta}k_{1\alpha}k_{2\beta}  \psi_{uudg}^{(53)}(1,2,3,4))
\right) \nonumber \\
  &&  \times \frac{\epsilon^{jkl}T^a_{li}}{2}\left[  u^{\dagger}_{i\uparrow}(1)
             \left(u^{\dagger}_{j\uparrow}(2)d^{\dagger}_{k\downarrow}(3)
            -d^{\dagger}_{j\uparrow}(2)u^{\dagger}_{k\downarrow}(3)\right)
	g^{a\dagger}_{\uparrow}(4)
	\right]\nonumber\\
&&+	\left( k_{1\perp}^-(\psi_{uudg}^{(54)}(1,2,3,4)
         + i\epsilon^{\alpha\beta}k_{2\alpha}k_{3\beta}  \psi_{uudg}^{(55)}(1,2,3,4))
\right. \nonumber \\
&&\left.+ k_{2\perp}^-(\psi_{uudg}^{(56)}(1,2,3,4)
         + i\epsilon^{\alpha\beta}k_{1\alpha}k_{3\beta}  \psi_{uudg}^{(57)}(1,2,3,4))
\right) \nonumber \\
  && \left. \times \frac{\epsilon^{jkl}T^a_{li}}{2}\left[  u^{\dagger}_{i\downarrow}(1)
             \left(u^{\dagger}_{j\uparrow}(2)d^{\dagger}_{k\uparrow}(3)
            -d^{\dagger}_{j\uparrow}(2)u^{\dagger}_{k\uparrow}(3)\right)
	g^{a\dagger}_{\uparrow}(4)
	\right]\right\}
         |0\rangle \ ,
\end{eqnarray}
where $\psi_{uudg}^{(54)}(1,2,3,4)=-\psi_{uudg}^{(54)}(1,3,2,4)$
and $\psi_{uudg}^{(55)}(1,2,3,4)=\psi_{uudg}^{(55)}(1,3,2,4)$.

Finally, for $l_z=-2$, the parton helicity must be 
$\lambda_{uud}=3/2$ and $\lambda_g=1$. We find 6 independent amplitudes
\begin{eqnarray}
  |P\uparrow\rangle_{uudg}^{l_z=-2} &=& \int d[1]d[2]d[3]d[4]
	\left( (k_{1\perp}^-)^2(\psi_{uudg}^{(58)}(1,2,3,4)
         + i\epsilon^{\alpha\beta}k_{2\alpha}k_{3\beta}  \psi_{uudg}^{(59)}(1,2,3,4))
\right. \nonumber \\
&&+(k_{2\perp}^-)^2(\psi_{uudg}^{(60)}(1,2,3,4)
         + i\epsilon^{\alpha\beta}k_{1\alpha}k_{3\beta}  \psi_{uudg}^{(61)}(1,2,3,4))
\nonumber \\
&&\left. + k_{1\perp}^-k_{2\perp}^-\psi_{uudg}^{(62)}(1,2,3,4)
	+k_{2\perp}^-k_{3\perp}^-\psi_{uudg}^{(63)}(1,2,3,4)\right) \nonumber \\
  &&  \times \frac{\epsilon^{jkl}T^a_{li}}{2}\left[  u^{\dagger}_{i\uparrow}(1)
             \left(u^{\dagger}_{j\uparrow}(2)d^{\dagger}_{k\uparrow}(3)
            -d^{\dagger}_{j\uparrow}(2)u^{\dagger}_{k\uparrow}(3)\right)
	g^{a\dagger}_{\uparrow}(4)
	\right]
         |0\rangle \ ,
\end{eqnarray}
where $\psi_{uudg}^{(58,63)}(1,2,3,4)=-\psi_{uudg}^{(58,63)}(1,3,2,4)$
and $\psi_{uudg}^{(59)}(1,2,3,4)=\psi_{uudg}^{(59)}(1,3,2,4)$.

The wave-function amplitudes for the other color structure
$(\epsilon^{ijl}T^a_{lk}+\epsilon^{ikl}T^a_{lj})/4$, 
can be defined similarly, except the
sign changes in the symmetric properties for some amplitudes.
We have in total of $63\times 2=126$ independent amplitudes
for the $uudg$ Fock component in the proton.

Note that the above construction is not unique, where we have first 
considered the correct flavor structure for the
three quarks, and then added all 
possible spin and color combinations. 
One can also start with a general spin structure 
for the three quarks, and then consider the isospin constraints. 
For example, for the total quark helicity $\lambda_{uud}=3/2$, the general spin
structure will be
\begin{equation}
\phi(1,2,3)\epsilon^{ijl}T^a_{lk}u_{i\uparrow}^\dagger (1)u_{j\uparrow}^\dagger (2)d_{k\uparrow}^\dagger (3)|0\rangle \ ,
\label{ma1}
\end{equation}
with the color coupling $\epsilon^{ijl}T^a_{lk}$. This color structure
says that indices $i$ and $j$ are antisymmetric, and 
the associated wave function amplitude $\phi(1,2,3)$ 
has $1 \leftrightarrow 2$ symmetry.
The isospin constraint indicates the following relations for the
three quark amplitude,
\begin{equation}
\phi(1,2,3)\epsilon^{ijl}T^a_{lk}\left(u_{i\uparrow}^\dagger (1)u_{j\uparrow}^\dagger (2)d_{k\uparrow}^\dagger (3)
+u_{i\uparrow}^\dagger (1)d_{j\uparrow}^\dagger (2)u_{k\uparrow}^\dagger (3)
+d_{i\uparrow}^\dagger (1)u_{j\uparrow}^\dagger (2)u_{k\uparrow}^\dagger (3)
\right)|0\rangle =0 \ .
\label{ma2}
\end{equation}
Applying the above relation to Eq.~(\ref{ma1}), 
and taking into account the $1\rightarrow 2$ symmetry for $\phi(1,2,3)$,
one has for the component,
\begin{equation}
\phi'(1,2,3)\epsilon^{ijl}T^a_{lk}u_{i\uparrow}^\dagger (1)\left(u_{j\uparrow}^\dagger (2)d_{k\uparrow}^\dagger (3)-
d_{j\uparrow}^\dagger (2)u_{k\uparrow}^\dagger (3)\right)|0\rangle \ ,
\end{equation}
where the isospin wave function for proton is explicit, and $\phi'(1,2,3)=\phi'(2,1,3)$. 
The similar analysis can be performed for the $\lambda_{uud}=\pm 1/2$ case. 
Working through all possibilities, we arrive at a
different set of wave-function amplitudes, 
which are essentially equivalent to the above construction. 
As a check, for every orbital angular momentum projection $l_z$, 
we find the same number of the independent amplitudes.

\section{wave-function amplitudes for the delta resonance}

In this section, we extend the above classification of the 
wave-function amplitudes to the baryon decuplet, assuming
again these are bound states.  
We consider one specific example, the delta resonance, 
and other baryon decuplets can be obtained 
by changing the flavor structure.
The distribution amplitudes for the $\Delta^{++}$ resonance have been
studied in Refs. \cite{Farrar:1989vz,Braun:1999te}.
$\Delta^{++}$ has two independent helicity states, $\Lambda=3/2$ and $1/2$,
and the other two helicity states $\Lambda=-3/2,\ -1/2$ can be obtained
by using the $\hat Y$ transformation Eq.~(\ref{jw}).

Here we classify only the three-quark amplitudes, and 
the total quark helicity is $\lambda_{uuu}=3/2,\ 1/2,\ -1/2,\ -3/2$.
We first consider $\Delta^{++}$ with helicity $\Lambda=3/2$, 
 The quark orbital angular momentum projection then has the following 
values $l_z=0,\ 1,\ 2,\ 3$ respectively. 
According to the method in the previous sections, we find 6
independent wave-function amplitudes,
\begin{eqnarray}
  |\Delta,\Lambda=3/2\rangle_{uuu}^{l_z=0} &=& \int d[1]d[2]d[3]\left( \psi_{uuu}^{(1)}(1,2,3)
         + i\epsilon^{\alpha\beta}k_{1\alpha}k_{2\beta}  \psi_{uuu}^{(2)}(1,2,3)\right) \nonumber \\
         &&  \times  \frac{\epsilon^{ijk}}{\sqrt{6}} u^{\dagger}_{i\uparrow}(1)
             u^{\dagger}_{j\uparrow}(2)u^{\dagger}_{k\uparrow}(3)|0\rangle \ , 
\end{eqnarray}
\begin{eqnarray}
  |\Delta,\Lambda=3/2\rangle_{uuu}^{l_z=1} &=& \int d[1]d[2]d[3]k_{1\perp}^+
           \psi_{uuu}^{(3)}(1,2,3) \nonumber\\
         &&  \times  \frac{\epsilon^{ijk}}{\sqrt{6}}  u^{\dagger}_{i\uparrow}(1)
            u^{\dagger}_{j\uparrow}(2)u^{\dagger}_{k\downarrow}(3)
         |0\rangle \ ,\\
  |\Delta,\Lambda=3/2\rangle_{uuu}^{l_z=2} &=& \int d[1]d[2]d[3]~\left(
k_{1\perp}^+k_{2\perp}^+\psi_{uuu}^{(4)}(1,2,3)
+k_{2\perp}^+k_{3\perp}^+\psi_{uuu}^{(5)}(1,2,3)\right)
 \nonumber \\
         &&  \times  \frac{\epsilon^{ijk}}{\sqrt{6}} u^{\dagger}_{i\uparrow}(1)
     u^{\dagger}_{j\downarrow}(2)u^{\dagger}_{k\downarrow}(3)
         |0\rangle \ , \\
  |\Delta,\Lambda=3/2\rangle_{uuu}^{l_z=3} &=& \int d[1]d[2]d[3]~
	k_{1\perp}^+k_{1\perp}^+k_{2\perp}^+
         \psi_{uuu}^{(6)}(1,2,3) \nonumber \\
         &&  \times  \frac{\epsilon^{ijk}}{\sqrt{6}} u^{\dagger}_{i\downarrow}(1)
             u^{\dagger}_{j\downarrow}(2)u^{\dagger}_{k\downarrow}(3)
         |0\rangle \ ,
\end{eqnarray}
where we have used symmetry to
reduce the number of independent amplitudes. 

For the helicity $\Lambda=1/2$ state of the $\Delta^{++}$ 
resonance, the classification is similar. The total quark helicity
is the same as the above, but the orbital angular projection $l_z$
can be $l_z=0,\ -1,\ 1,\ 2$.
As a result, the three quark Fock component for $\Delta^{++}(\Lambda=1/2)$ 
has the following 5 independent wave-function amplitudes,
\begin{eqnarray}
  |\Delta,\Lambda=1/2\rangle_{uuu}^{l_z=0} &=& \int d[1]d[2]d[3]\left( \psi_{uuu}^{(1)}(1,2,3)
         + i\epsilon^{\alpha\beta}k_{1\alpha}k_{2\beta}  \psi_{uuu}^{(2)}(1,2,3)\right) \nonumber \\
         &&  \times  \frac{\epsilon^{ijk}}{\sqrt{6}} u^{\dagger}_{i\uparrow}(1)
             u^{\dagger}_{j\uparrow}(2)u^{\dagger}_{k\downarrow}(3)|0\rangle \ , \\
  |\Delta,\Lambda=1/2\rangle_{uuu}^{l_z=1} &=& \int d[1]d[2]d[3]k_{2\perp}^+
           \psi_{uuu}^{(3)}(1,2,3) \nonumber\\
         &&  \times  \frac{\epsilon^{ijk}}{\sqrt{6}}  u^{\dagger}_{i\uparrow}(1)
            u^{\dagger}_{j\downarrow}(2)u^{\dagger}_{k\downarrow}(3)
         |0\rangle \ ,\\
  |\Delta,\Lambda=1/2\rangle_{uuu}^{l_z=-1} &=& \int d[1]d[2]d[3]~
k_{2\perp}^-\psi_{uuu}^{(4)}(1,2,3)
 \nonumber \\
         &&  \times  \frac{\epsilon^{ijk}}{\sqrt{6}} u^{\dagger}_{i\uparrow}(1)
     u^{\dagger}_{j\uparrow}(2)u^{\dagger}_{k\uparrow}(3)
         |0\rangle \ , \\
  |\Delta,\Lambda=1/2\rangle_{uuu}^{l_z=2} &=& \int d[1]d[2]d[3]~
	k_{1\perp}^+k_{2\perp}^+
         \psi_{uuu}^{(5)}(1,2,3) \nonumber \\
         &&  \times  \frac{\epsilon^{ijk}}{\sqrt{6}} u^{\dagger}_{i\downarrow}(1)
             u^{\dagger}_{j\downarrow}(2)u^{\dagger}_{k\downarrow}(3)
         |0\rangle \ .
\end{eqnarray}
We will not go beyond the three-quark Fock components, although those
can be classified similarly.

Since the flavor structure for the baryon decuplet is completely
symmetric, the light-cone expansion for the 
other states in the decuplet can be easily written
down from the above results, apart from 
that the flavor structure need to be replaced accordingly.
For example, the $\Delta^{+}$ resonance has the symmetric 
flavor structure in the form of $(uud+udu+duu)/\sqrt{3}$. 
Substituting this into the above equations, the light-cone
expansion of $\Delta^{+}$ for the three quark Fock component
can be obtained. 
These amplitudes, together with those for
the proton, are needed to calculate the proton-delta 
transition form factors in the asymptotic limit of QCD 
\cite{Brodsky:1981sx,Carlson:1986mm,Carlson:1987zs}.

\section{Asymptotic Scaling of Wave-function amplitudes}

One of the important applications of the light-cone wave-function
amplitudes is to calculate hard exclusive processes.
The relative importance of a particular amplitude
in a process can be determined from its scaling behavior
when the parton transverse-momenta become large. 
In this section, we follow the discussion in Ref. \cite{Ji:2003fw}
and derive a general power counting rule for the light-cone
amplitudes of any Fock components of a hadron state. 
As examples, we consider the asymptotic scaling
of some amplitudes defined above for the $\pi^+$ and proton. 
 
Let $\psi_{n}(x_i,k_i,l_{zi})$ be a general amplitude describing 
a $n$ partons Fock component of a hadron state with orbital angular 
momentum projection of $l_z=\sum_il_{zi}$.
To find the asymptotic behavior of $\psi_{n}(x_i,k_i,l_{zi})$
in the limit that all transverse momenta are uniformly large, 
we consider the matrix element of a corresponding quark-gluon 
operator between the QCD vacuum and the hadron state
\begin{equation}
\langle 0|\phi_{\mu_1}(\xi_1)....\phi_{\mu_n}(\xi_n)|P\Lambda\rangle \ , 
\end{equation}
where $\phi$ are parton fields such as the ``good" (+) components of
quark fields or $F^{+\alpha}$ of gluon fields, and
$\mu_i$ are Dirac and transverse coordinate indices when appropriate. 
All spacetime coordinates $\xi_i$ are at equal 
light-cone time, $\xi_i^+=0$. Fourier-transforming with respect to all the spatial 
coordinates ($\xi^-_i$,$\xi_{i\perp}$), we find the matrix element in 
the momentum space, $
\langle 0|\phi_{\mu_1}(k_1)....\phi_{\mu_{n-1}}(k_{n-1})
       \phi_{\mu_n}(0)|p\Lambda\rangle \equiv \psi_{\mu_1,...,\mu_n}(k_1,...,k_{n-1}) $,
here we have just shown $n-1$ parton momenta because of 
the overall momentum conservation. The matrix element
can be written as a sum of terms involving projection
operator $\Gamma^{A}_{\mu_1...\mu_n}(k_{i\perp})$
multiplied by scalar amplitude $\psi_{nA}(x_i,k_{i\perp},l_{zi})$: 
\begin{eqnarray}
\langle 0|\phi_{\mu_1}(k_1)....\phi_{\mu_{n-1}}(k_{n-1})
       \phi_{\mu_n}(0)|p\Lambda\rangle  
  && \equiv  \psi_{\mu_1,...,\mu_n}(k_1,...,k_{n-1}) \nonumber \\
  && = \sum_{A} \Gamma^{A}_{\mu_1...\mu_n}(k_{i\perp})
          \psi_{n}^{(A)}(x_i,k_{i\perp},l_{zi}) \ , 
\end{eqnarray}
where the projection operator $\Gamma^A$ contains 
Dirac matrices and is a polynomial of order $l_z$ 
in parton momenta. For example, the two
quark matrix element of the pion can be written as \cite{Burkardt:2002uc},
\begin{eqnarray}
     &&   \langle 0|\overline{d}_{+\mu}(0)u_{+\nu}(x,k_\perp) 
    |\pi^+(P)\rangle \nonumber \\
   && = (\gamma_5\not\! P)_{\nu\mu} \psi_{u\overline{d}}^{(1)}(x,k_\perp,l_z=0)
            + (\gamma_5\sigma^{-\alpha})_{\nu\mu}P^+ k_{\perp \alpha} 
        \psi_{u\overline{d}}^{(2)}(x,k_\perp,l_z=1) \ ,
\end{eqnarray}
where the projection operators are shown manifestly. More examples for the proton
matrix elements can be found in Ref. \cite{Ji:2002xn}. 

The matrix element of our interest is, in fact, a 
Bethe-Salpeter amplitude projected onto the light cone. 
One can write down formally a Bethe-Salpeter equation 
which includes mixing contributions from other light-cone 
matrix elements.
In the limit of large transverse momentum $k_{i\perp}$, 
the Bethe-Salpeter kernels can be calculated 
in perturbative QCD because of asymptotic freedom. 
In the lowest order, the kernels consist of a minimal 
number of gluon and quark exchanges linking the active 
partons. For the lowest Fock components of the
pion wave function, one gluon exchange is needed to get a large 
transverse momentum for both quarks \cite{Lepage:1980fj}.
As we shall see, asymptotic behavior of the wave function 
amplitudes depends on just the mass dimension of the kernels.

Schematically, we have the following equation for the light-cone
amplitudes,
\begin{eqnarray}
&& \psi_{\alpha_1,...,\alpha_n}(k_1,...,k_{n-1}) 
= \sum_A \Gamma^A_{\alpha_1...\alpha_n}(k_{i\perp}) 
      \psi_{n}^A(x_i, k_{i\perp}, l_{zi}) 
   \nonumber \\
&= & \sum_{n',\beta_1,...,\beta_{n'}}\int d^4q_1 ... d^4q_{n'-1}
       H_{\alpha_1...,\alpha_n,\beta_1,...,\beta_{n'}}(q_i,
      k_i) \psi_{\beta_1,...,\beta_{n'}}(q_1,...,q_{n'-1}) 
\label{bs}
\end{eqnarray}
where $H_{\alpha_1,...,\alpha_n,\beta_1,...,\beta_{n'}}$ are
the Bethe-Salpeter kernels multiplied by the parton propagators. 
When the parton transverse momenta are uniformly large, the kernels
can be approximated by a sum of perturbative diagrams. 
The leading contribution 
to the amplitudes on the left can be obtained by
iterating the above equation, assuming the amplitudes under
the integration sign contain no hard components. As such,
the integrations over $q_{i\perp}$ can be cut-off at a scale $\mu$
where $k_{\perp}>\!\!>\mu>\!\!>\Lambda_{\rm QCD}$, 
and the $q_i$ dependence in $H$ can be expanded in Taylor series. 
In order to produce a contribution to $\psi_n^{(A)}
(x_i,k_{i\perp},l_{zi})$, the hard kernels must contain 
the projection operator $\Gamma^A_{\alpha_1...\alpha_n}(k_1,...,k_{n-1})$. 
Hence we write
\begin{eqnarray}
 && H_{\alpha_1...,\alpha_n,\beta_1,...,\beta_n'}(q_i,k_i)  \nonumber \\
  &=&  \sum_{A,B} \Gamma^A_{\alpha_1...\alpha_n}(k_{i\perp}) 
      H_{AB}(x_i, k_{i\perp}, y_i)
       \Gamma^B_{\beta_1...\beta_{n'}}(q_{i\perp}) \ , 
\end{eqnarray}
where $\Gamma^B_{\beta_1...\beta_{n'}}(q_{i\perp})$ is again a projection 
operator and $H_{AB}(x_i,k_i,y_i)$ are scalar functions of the
transverse momenta $k_{i\perp}$ invariants. Substituting the above into Eq.(\ref{bs}) 
and integrating over $q_i^-$, 
\begin{eqnarray}
&& \psi^{(A)}_n(x_i,k_{i\perp}, l_{zi})
   \nonumber \\
&= & \sum_{B,\beta_i} \int dy_1...dy_{n'-1} H_{AB}(x_i,k_i,y_i) \int d^2q_{1\perp} ... 
    d^2q_{(n'-1)\perp}
    \Gamma^B_{\beta_1...\beta_{n'}}(q_{i\perp})
          \psi_{\beta_1,...,\beta_{n'}}(y_i,q_i)  \nonumber \\
&= & \sum_{B,\beta_i,A'}\int dy_1...dy_{n'-1} H_{AB}(x_i,k_i,y_i) \int d^2q_{1\perp} ... 
    d^2q_{(n'-1)\perp}
    \Gamma^B_{\beta_1...\beta_{n'}}(q_{i\perp}) \nonumber \\ && 
      \times  \Gamma^{A'}_{\beta_1...\beta_{n'}}(q_{i\perp})
       \psi_{n'}^{(A')}(y_i,q_{i\perp},l_{zi}') \ ,  
\label{e7}
\end{eqnarray}
where the integrations over $q_{i\perp}$ are non-zero
only when the angular momentum content of $\Gamma^B$
and $\Gamma^{A'}$ is the same. Now the large momenta $k_{i\perp}$ 
are entirely isolated in $H_{AB}$ which does not depend on 
any soft scale. The asymptotic behavior of $\psi_n^{(A)}(k_{i\perp})$ is 
determined by the mass dimension of $H_{AB}$, which can be obtained,
in principle, by working through one of the simplest perturbative diagrams.

A much simpler way to proceed is to use light-cone power 
counting in which the longitudinal mass dimension, such as 
$P^+$, can be ignored because of the boost
invariance of the above equation along the $z$ direction.
We just need to focus on the transverse dimensions. 
The mass dimension of the light-cone wave function amplitude 
$\psi_n$ can be determined as follows:
Assume the hadron state is normalized relativistically
$\langle P|P'\rangle = 2E(2\pi)^3\delta^3(\vec{P}'-\vec{P})$, 
$|P\rangle$ has mass dimension $-1$. Likewise, the parton creation 
operator $a_i^\dagger$ has mass dimension $-1$. Given these, the mass dimension
of $\psi_n$ is $-(n+|l_z|-1)$ according to Eq. (\ref{gs}). 
The mass dimension of $\psi_{n(ij)}$ term in (\ref{gs}), however, 
is $-(n+|l_z|+1)$ which can be accounted for by the previous formula with 
an effective angular momentum projection $|l_z|+2$. 

Now considering Eq. (\ref{e7}), since the mass dimension of the amplitude
$\psi^{(A)}_n(x_i,k_{i\perp}, l_{zi})$ is $-(n+|l_z|-1)$,
that of $\Gamma^B\Gamma^{A'}$ is $2|l_z'|$, 
and the integration measure $2(n'-1)$, a
balance of the mass dimension yields
$[H_{AB}] = -(n-1+|l_z|)-(n'-1+|l_z'|)$. Therefore, the leading behavior 
of the wave function amplitude goes as  \cite{Ji:2003fw}
\begin{equation}
   \psi^{(A)}_n(x_i,k_{i\perp}, l_{zi}) \sim \frac{1}
        {(k_\perp^2)^{[n+|l_z|+{\rm min}(n'+|l_z'|)]/2-1}} \ , 
\label{scaling}
\end{equation}
which is determined by a mixing amplitude with smallest
$n'+|l_z'|$. Since the wave function amplitude has mass dimension of 
$-(n+|l_z|-1)$, the coefficient of the asymptotic form must have a soft mass
dimension $\Lambda_{\rm QCD}^{{\rm min}(n'+|l_z'|)-1}$. 

We have the following selection rules for amplitude mixings.
First of all, because
of angular momentum conservation, wave function amplitudes 
belonging to different hadron helicity states do not mix. Second, because
of the vector coupling in QCD, the quark helicity in a hard process 
does not change. Therefore, the pion amplitude $\psi^{(2)}_{u\bar d}$ does not 
mix with $\psi^{(1)}_{u\bar d}$ because the total quark helicity differs. 
An example of the nontrivial amplitude mixing is between the pion's 
two-quark-one-gluon and two-quark amplitudes. 

The power counting rule Eq.~(\ref{scaling}) can be used
to predict the scaling behaviors for all the 
light-cone wave function amplitudes we have written down in the
above for the mesons and baryons. We will not go into much
details about these predictions, rather we consider some examples
for the $\pi^+$ and proton. 
According to Eq. (\ref{scaling}), the scaling
behaviors for the two-parton light-cone 
amplitudes of $\pi^+$ are
\begin{equation}
\psi_{u\overline d}^{(1)}(1,2)\sim 1/k_\perp^2,~~~~
\psi_{u\overline d}^{(2)}(1,2)\sim 1/k_\perp^4 \ .
\end{equation}
The $u\bar d g$ Fock amplitudes have the following
scaling,
\begin{equation}
\psi_{u\overline d g}^{(1,3,4,5,6)}(1,2,3)\sim 1/k_\perp^4,~~
\psi_{u\overline d g}^{(2,7,8,9)}(1,2,3)\sim 1/k_\perp^6, \
\end{equation}
where the mixings with the two-parton components give the dominant
contribution at large $k_\perp$.

For the three quark Fock component of the proton, 
we have the following
scaling behaviors for the light-cone amplitudes,
\begin{equation}
\psi_{uud}^{(1)}\sim 1/k_\perp^4,~~
\psi_{uud}^{(2,3,4,5)}\sim 1/k_\perp^6,~~
\psi_{uud}^{(6)}\sim 1/k_\perp^8.
\end{equation}
Here, the scaling behaviors of $\psi_{uud}^{(1,3,4,5,6)}$ at large 
$k_\perp$ are determined by self mixings, 
while that of $\psi_{uud}^{(2)}$ is determined by mixing with
$\psi_{uud}^{(1)}$.

\section{summary and conclusion}

Following Ref. \cite{Ji:2003fw}, we studied in this paper 
how to classify the independent 
wave-function amplitudes for a hadron state. 
We discussed in detail how the spin, flavor (for quark) and 
color of the partons are systematically coupled. 
We have found the these amplitudes for  
pion and proton up to and including four partons. 
We also worked out the leading light-cone wave amplitudes 
for the $\Delta$ resonance and the $\rho$ meson.

A general power counting rule for the light-cone wave function
amplitude has been derived based on perturbative QCD \cite{Ji:2003fw}. 
Using this rule, we have predicted the asymptotic scaling behavior of 
a number of amplitudes for $\pi^+$ and the proton.
This general power counting rule can be used as a constraint 
in modeling the light-cone wave function
amplitudes.

Many applications can be made based on
the formalism presented here. One example is
the generalized power counting rule for high energy
exclusive processes \cite{Ji:2003fw}, including processes
involving nonzero parton orbital angular momentum and
hadron helicity flip. A number of processes have been
briefly discussed in \cite{Ji:2003fw}. A more
detailed discussion of the generalized counting rule 
will be presented elsewhere. In a different direction,
one can also parameterize the light-cone 
wave function amplitudes and fit them to many relevant
experiment data, such as the elastic form factors,
parton distributions, and generalized parton distributions.

X. J. and F. Y. were supported by the U. S. Department of Energy via 
grant DE-FG02-93ER-40762. J.P.M. was supported by National Natural 
Science Foundation of P.R. China through grand No.19925520. 

\bibliography{protonref}

\end{document}